\documentclass[printer]{aa}
\usepackage{graphicx}
\usepackage{txfonts}
\def\lesssim{\mathrel{\hbox{\rlap{\hbox{\lower4pt\hbox{$\sim$}}}\hbox{$<$}}}}
\def\gtrsim{\mathrel{\hbox{\rlap{\hbox{\lower4pt\hbox{$\sim$}}}\hbox{$>$}}}}
\newcommand{\mincir}{\raise -2.truept\hbox{\rlap{\hbox{$\sim$}}\raise5.truept
\hbox{$<$}\ }}
\newcommand{\magcir}{\raise -2.truept\hbox{\rlap{\hbox{$\sim$}}\raise5.truept
\hbox{$>$}\ }}
\newcommand{\siml}{\raise -2.truept\hbox{\rlap{\hbox{$\sim$}}\raise5.truept
\hbox{$<$}\ }}
\newcommand{\simg}{\raise -2.truept\hbox{\rlap{\hbox{$\sim$}}\raise5.truept
\hbox{$>$}\ }}
\newcommand{\be}{\begin{equation}}
\newcommand{\ee}{\end{equation}}
\newcommand{\ba}{\begin{eqnarray}}
\newcommand{\ea}{\end{eqnarray}}
\newcommand {\kpc} {$h_{70}^{-1}$ kpc$\;$}
\newcommand {\kpcc} {$h_{70}^{-1}$ kpc}
\newcommand {\h} {$h_{70}^{-1}$ Mpc$\;$}
\newcommand {\hh} {$h_{70}^{-1}$ Mpc}
\newcommand {\hhh} {\;h_{70}^{-1} \mathrm{Mpc}}
\newcommand {\ks} {km~s$^{-1} \;$}
\newcommand {\kss} {km~s$^{-1}$}

\newcommand {\mqua} {$\times 10^{14}\;h_{70}^{-1}\;M_{\odot} \;$}
\newcommand {\mquaa} {$\times 10^{14}\;h_{70}^{-1}\;M_{\odot}$}
\newcommand {\mqui} {$\times 10^{15}\;h_{70}^{-1}\;M_{\odot} \;$}
\newcommand {\mquii} {$\times 10^{15}\;h_{70}^{-1}\;M_{\odot}$}

\newcommand {\mll} {$h_{70}\;M_{\odot}/L_{\odot}$}

\newcommand{\degree}{\ensuremath{\mathrm{^\circ}}}
\newcommand{\arcm}{\ensuremath{\mathrm{^\prime}\;}}
\newcommand{\arcs}{\ensuremath{\arcmm\hskip -0.1em\arcmm \;}}
\newcommand{\arcmm}{\ensuremath{\mathrm{^\prime}}}
\newcommand{\arcss}{\ensuremath{\arcmm\hskip -0.1em\arcmm}}
\newcommand{\dotarcs}{\,\rlap{\hbox{$\mathrm{^\prime\hskip-0.1em^\prime}$}}{\hbox{$.$}}\,}

\newcommand{\dotsec}{\,\rlap{\hbox{$\mathrm{^s}$}}{\hbox{$.$}}\,}
\begin{document}
   \title{Internal dynamics of the radio halo cluster Abell~773:
\\ a multiwavelength analysis}
%
\author{R. Barrena\inst{1}
\and W. Boschin\inst{2,3} 
\and M. Girardi\inst{3,4}
\and M. Spolaor\inst{3,5}
}
   \offprints{R. Barrena; e.mail: rbarrena@iac.es}

\institute{
Instituto de Astrofisica de Canarias, C/Via Lactea s/n, E-38205 La Laguna, Tenerife, Canary Islands, Spain\\
\and 
Fundaci\'on Galileo Galilei - INAF, C/Alvarez de Abreu 70, E-38700 Santa Cruz de La Palma, Canary Islands, Spain\\
\and
Dipartimento di Astronomia, Universit\`{a} degli Studi di Trieste, via Tiepolo 11, I-34131 Trieste, Italy\\
\and
INAF -- Osservatorio Astronomico di Trieste, via Tiepolo 11, I-34131  Trieste, Italy\\
\and
Centre for Astrophysics \& Supercomputing, Swinburne University, Hawthorn, VIC 3122, Australia\\
}

\date{Received / accepted }

\abstract{} {To conduct an intensive study of the rich, X-ray luminous
galaxy cluster Abell 773 at $z=0.22$ containing a diffuse radio halo
to determine its dynamical status.}{Our analysis is based on new
spectroscopic data obtained at the TNG telescope for 107 galaxies, 37
spectra recovered from the CFHT archive and new photometric data
obtained at the Isaac Newton Telescope.  We use statistical tools to
select 100 cluster members (out to $\sim$1.8 \h from the cluster
centre), to analyse the kinematics of cluster galaxies and to
determine the cluster structure. Our analysis is also performed by
using X-ray data stored in the Chandra archive.}  {The 2D distribution
shows two significant peaks separated by $\sim$2\arcm in the EW
direction with the main western one closely located at the position of
the two dominant galaxies and the X-ray peak.  The velocity
distribution of cluster galaxies shows two peaks at ${\rm v}\sim$65000
and $\sim$67500 \kss, corresponding to the velocities of the two
dominant galaxies.  The low velocity structure has a high velocity
dispersion -- $\sigma_{\rm v}=800-1100$ \ks -- and its galaxies are
centred on the western 2D peak.  The high velocity structure has
intermediate velocity dispersion -- $\sigma_{\rm v}\sim 500$ \ks --
and is characterized by a complex 2D structure with a component
centred on the western 2D peak and a component centred on the eastern
2D peak, these components probably being two independent groups.  We
estimate a cluster mass within 1 \h of 6--11 \mquaa. Our analysis of
Chandra data shows the presence of two very close peaks in the core
and the elongation of the X-ray emission in the NEE--SWW
direction.  }{Our results suggest we are looking at probably two
groups in an advanced stage of merging with a main cluster and having
an impact velocity $\Delta {{\rm v}_{\rm rf}} \sim 2300$ \kss. In
particular, the radio halo seems to be related to the merger of the
eastern group.}


\authorrunning{Barrena et al.}
\titlerunning{Internal dynamics of A773} 
\maketitle
%

\section{Introduction}

Clusters of galaxies are recognized to be not simple relaxed
structures, but rather as evolving via merging processes in a
hierarchical fashion from poor groups to rich clusters. Much progress
has been made in recent years in the observations of the signatures of
merging processes (see Feretti et al. \cite{fer02b} for a general
review). The presence of substructure, which is indicative of a
cluster at an early phase of the process of dynamical relaxation or of
secondary infall of clumps into already virialized clusters occurs in
about 50\% of clusters as shown by optical and X-ray data (e.g.\
Geller \& Beers \cite{gel82}; Mohr et al. \cite{moh96}; Girardi et
al. \cite{gir97}; Kriessler \& Beers \cite{kri97}; Jones \& Forman
\cite{jon99}; Schuecker et al. \cite{sch01}; Burgett et
al. \cite{bur04}).

A very interesting aspect of these investigations is the possible
connection of cluster mergers with the presence of extended diffuse
radio sources, halos and relics. They are rare, large (up to
$\sim1$--2 \hh), amorphous cluster sources of uncertain origin and
generally steep radio spectra (Hanisch \cite{han82}; see also
Giovannini \& Feretti \cite{gio02} for a recent review).  They appear
to be associated with very rich clusters that have undergone recent
mergers, and it has therefore been suggested by various authors that
cluster halos/relics are related to recent merger activity (e.g.\
Tribble \cite{tri93}; Burns et al. \cite{bur94}; Feretti
\cite{fer99}).

The synchrotron radio emission of halos and relics demonstrates the
existence of large scale cluster magnetic fields, of the order of
0.1--1 $\mu$G, and of widespread relativistic particles of energy
density 10$^{-14}$ -- 10$^{-13}$ erg cm$^{-3}$.  The difficulty in
explaining radio halos arises from the combination of their large size
and the short synchrotron lifetime of relativistic electrons.  The
expected diffusion velocity of the electron population is of the order
of the Alfven speed ($\sim$100 \kss) making it difficult for the
electrons to diffuse over a megaparsec-scale region within their
radiative lifetime. Therefore, one needs a mechanism by which the
relativistic electron population can be transported over large
distances in a short time, or a mechanism by which the local electron
population is reaccelerated and the local magnetic fields are
amplified over an extended region.  The cluster--cluster merger can
potentially supply both mechanisms (e.g.\ Giovannini et
al. \cite{gio93}; Burns et al. \cite{bur94}; R\"ottgering et
al. \cite{rot94}; see also Feretti et al. \cite{fer02a}; Sarazin
\cite{sar02} for reviews). However, the question is still debated
since the diffuse radio sources are quite uncommon and only recently
we have been able to study these phenomena on the basis of sufficient
statistics (a few dozen clusters up to $z\sim0.3$, e.g.\ Giovannini et
al. \cite{gio99}; see also Giovannini \& Feretti \cite{gio02}; Feretti
\cite{fer05}).

Growing evidence for the connection between diffuse radio emission and
cluster merging is based on X-ray data (e.g.\ B\"ohringer \& Schuecker
\cite{boh02}; Buote \cite{buo02}). Studies based on a large number of
clusters have found a significant relation between the radio and the
X-ray surface brightness (Govoni et al. \cite{gov01a}, \cite{gov01b})
and between the presence of radio halos/relics and irregular and
bimodal X-ray surface brightness distribution (Schuecker et
al. \cite{sch01}).  New unprecedent insights into merging processes in
radio clusters are offered by Chandra and XMM--Newton observations
(e.g.\ Markevitch \& Vikhlinin \cite{mar01}; Markevitch et
al. \cite{mar02}; Fujita et al. \cite{fuj04}; Henry et
al. \cite{hen04}; Kempner \& David \cite{kem04}).

Optical data are a powerful way to investigate the presence and the
dynamics of cluster mergers, too (e.g.\ Girardi \& Biviano
\cite{gir02}).  The spatial and kinematic analysis of member galaxies
allow us to detect and measure the amount of substructure, to identify
and analyse possible pre-merging clumps or merger remnants.  This
optical information is complementary to X-ray information since
galaxies and ICM react on different time scales during a merger (see,
for example, numerical simulations by Roettiger et al. \cite{roe97}).
Unfortunately, to date optical data are lacking or are poorly
exploited.  The sparse literature contains a few individual clusters
(e.g.\ Colless \& Dunn \cite{col96}; G\'omez et al. \cite{gom00};
Barrena et al. \cite{bar02}; Mercurio et al. \cite{mer03}; Boschin et
al. \cite{bos04}; Boschin et al. \cite{bos06}; Girardi et
al. \cite{gir06}). We have conducted an intensive study of Abell 773
(hereafter A773), which has a radio halo (Giovannini et
al. \cite{gio99}, see also Kempner \& Sarazin \cite{kem01}).

A773 is a rich, X-ray luminous and hot cluster at $z\sim 0.22$: Abell
richness class = 2 (Abell et al. \cite{abe89}),
$L_\mathrm{X}$(0.1--2.4 keV)$=12.52\times 10^{44} \ h_{50}^{-2}$ erg\
s$^{-1}$ (Ebeling et al. \cite{ebe96}), $T_\mathrm{X}\sim7$--9 keV
(Allen \& Fabian \cite{all98}; Rizza et al. \cite{riz98}; White
\cite{whi00}; Saunders et al. \cite{sau03}; Govoni et
al. \cite{gov04}; Ota \& Mitsuda \cite{ota04}).  This cluster has a
centre that is extremely optically rich and contains almost only
early-type galaxies (Dahle et al. \cite{dah02}).  A lot of evidence
points towards the young dynamical status of A773.  This cluster is
elongated along the NE--SW direction and has a great amount of
substructure in X-ray emission, as shown by ROSAT HRI data (Pierre \&
Starck \cite{pie98}; Rizza et al. \cite{riz98}; Govoni et
al. \cite{gov01b}).  There is a strong peak in the mass map, offset to
the southwest from the optical centre by about 1\arcm and the mass
appears to be elongated in a different way with respect to the light
and number density distribution (Dahle et al. \cite{dah02}).  The
optical DSS image clearly shows two galaxy substructures, one in the
centre of the X-ray emission and another at the eastern region (Govoni
et al. \cite{gov04}).

To date, few spectroscopic data have been reported in the literature.
Crawford et al. (\cite{cra95}) measured the redshift for the two
brightest, equally dominant galaxies ($z\sim 0.22$).  A few redshifts
for radio galaxies in the cluster region are given by Morrison et
al. (\cite{mor03}). We have recently carried out spectroscopic
observations with the TNG giving new redshift data for 107 galaxies in
the field of A773, as well as photometric observations at the INT.  We
recover additional spectroscopic information from reduction of CFHT
archive data.  Our present analysis is based on these optical data and
X-ray Chandra archival data.

This paper is organized as follows.  We present the new optical data
in Sect.~2 and the relevant analyses in Sect.~3.  Our analysis of
Chandra X-ray data is shown in Section~4. We discuss the dynamical
status of A773 in Sect.~5 and summarize our results in Section~6.

Unless otherwise stated, we give errors at the 68\% confidence level
(hereafter c.l.).  Throughout the paper, we assume a flat cosmology
with $\Omega_{\rm m}=0.3$, $\Omega_{\Lambda}=0.7$ and $H_0=70$
$h_{70}$ \ks Mpc$^{-1}$. For this cosmological model, 1\arcm
corresponds to 213 \kpc at the cluster redshift.

\section{Data sample}

\subsection{Spectroscopy}

\begin{figure}
\centering
\resizebox{\hsize}{!}{\includegraphics{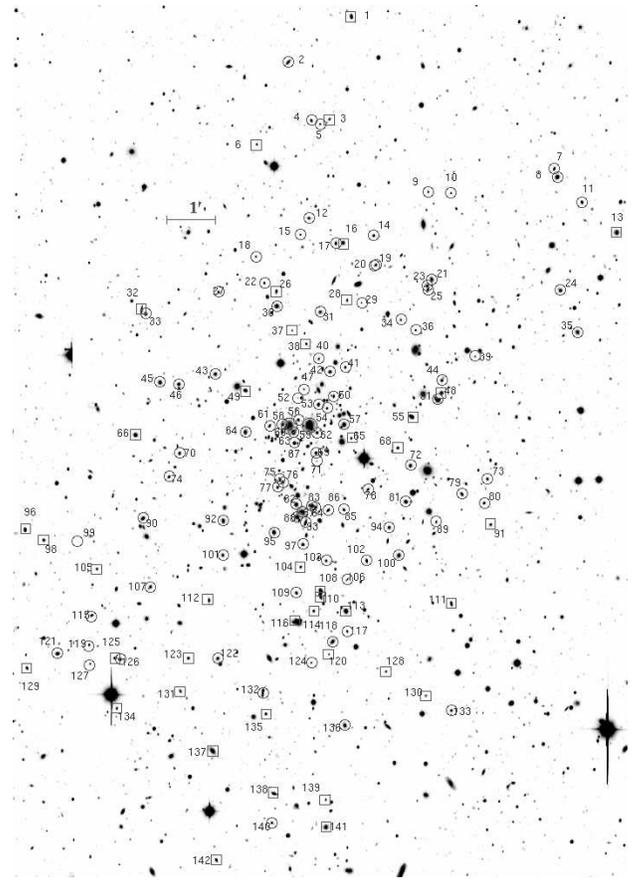}}
\caption{$R$-band image of A773 (west at the top and north to the
left) taken with the WFC camera of the INT.  Galaxies with successful
velocity measurements are labelled as in Table~\ref{tab1}.  Circles
and boxes indicate cluster members and non-member galaxies,
respectively.}
\label{figimage}
\end{figure}

\begin{figure}
\centering
\resizebox{\hsize}{!}{\includegraphics{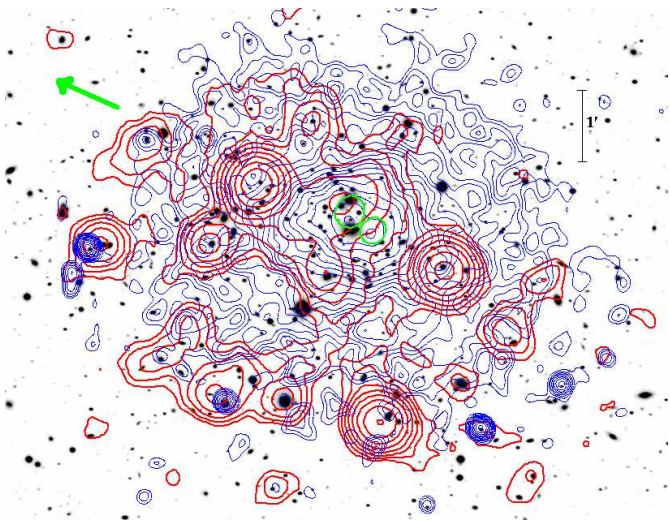}} 
\caption{$R$-band image of the cluster A773 with, superimposed, the
contour levels of the Chandra archival image ID 5006 (blue; photons in
the energy range 0.3--7 keV) and the contour levels of a VLA radio
image (red) at 1.4 GHz (Govoni et al. \cite{gov01b}).  Green ellipses
identify structures detected by Wavdetect.  The direction of A782 is
also shown (see Sect.~5.4).  North is at the top and east to the
left.}
\label{figisofote}

\end{figure}

Multi-object spectroscopic observations of A773 were carried out at
the TNG in December 2003 during the programme of proposal
AOT8/CAT--G6. We used DOLORES/MOS with the LR--B Grism 1, giving a
dispersion of 187 \AA/mm, and the Loral $2048\times2048$ CCD (pixel
size 15 $\mu$m). This combination of grating and detector results in a
dispersion of 2.8 \AA/pix. We have taken four MOS masks for a total of
137 slits. We acquired two exposures of 1800 s for one mask and three
exposures of 1800 s for three masks. Wavelength calibration was
performed using helium--argon lamps.

Reduction of spectroscopic data was carried out with the IRAF package.
\footnote{IRAF is distributed by the National Optical Astronomy
Observatories, which are operated by the Association of Universities
for Research in Astronomy, Inc., under cooperative agreement with the
National Science Foundation.}

Radial velocities were determined using the cross-correlation
technique (Tonry \& Davis \cite{ton79}) implemented in the RVSAO
package (developed at the Smithsonian Astrophysical Observatory
Telescope Data Center).  Each spectrum was correlated against six
templates for a variety of galaxy spectral types: E, S0, Sa, Sb, Sc
and Ir (Kennicutt \cite{ken92}).  The template producing the highest
value of $\cal R$, i.e.\ the parameter given by RVSAO and related to
the signal-to-noise of the correlation peak, was chosen.  Moreover,
all the spectra and their best correlation functions were examined
visually to verify the redshift determination.  In two cases (IDs~66
and 129; see Table~\ref{tab1}) we took the EMSAO redshift as a
reliable estimate of the redshift.  We obtained redshifts for 107
galaxies.

For six galaxies we obtained two redshift determinations of similar
quality.  This allows us to obtain a more rigorous estimate for the
redshift errors since the nominal errors as given by the
cross-correlation are known to be smaller than the true errors (e.g.\
Malumuth et al.  \cite{mal92}; Bardelli et al. \cite{bar94}; Ellingson
\& Yee \cite{ell94}; Quintana et al. \cite{qui00}).  For the six
galaxies having two redshift determinations we fit the first
determination vs.\ the second one by using a straight line and
considering errors in both coordinates (e.g.\ Press et
al. \cite{pre92}). The fitted line agrees with the one-to-one
relation, but, when using the nominal cross-correlation errors, the
small value of $\chi^2$ probability indicates a poor fit, suggesting
the errors are underestimated.  Only when nominal errors are
multiplied by a factor of $\sim$1.5 can the observed scatter be
justified. We therefore assume hereafter that true errors are larger
than nominal cross-correlation errors by a factor 1.5. For the six
galaxies we used the average of the two redshift determinations and
the corresponding error.

We added to our data observations stored in the CFHT archive (proposal
ID: 01AF37): 1 MOS mask of three 1800 s exposures with 52 slits. We
reduced the spectra with the same procedure adopted for the TNG
observations. We succeeded in obtaining spectra for 37 galaxies, two
of which are in common with TNG galaxies. The comparison of the two
estimates agree well within the errors ($cz_{\rm CFHT}=44170\pm199$
vs.\ $cz_{\rm TNG}=44195\pm94$ \kss, $cz_{\rm CFHT}=64304\pm113$ vs.
$cz_{\rm TNG}=64260\pm46$ \kss), allowing us to combine the data.  For
the two galaxies in common we used the average of the two redshift
determinations and the corresponding error.

Our spectroscopic catalogue consists of 142 galaxies in a region of
18\arcmm$\times$13\arcm centred on the position of the two close
dominant galaxies (IDs~59 and 60, hereafter D1 and D2; see
Table~\ref{tab1}).  The median $S/N$ and error on $cz$ are 12 and 72
\kss, respectively.

We use a conservative approach leading to a sparse spectral
classification (75\% of the sample, see Table~\ref{tab1}).  We follow
the classification of Dressler et al. (\cite{dre99}; see also
Poggianti et al. \cite{pog99}).  We define ``e''-type galaxies as
those showing active star formation indicated by the presence of
[OII], and, in particular, ``e(b)'' galaxies as those showing an
equivalent width EW([OII]) of $\le-40$ \AA $\ $ (probably starburst
galaxies).  Of galaxies having $S/N\gtrsim 10\,\,$ we define as
``k''-type those having EW(H$\delta$)$<$3 \AA $\,$ and no
emission lines (probably passive galaxies); ``k+a''- and ``a+k''-type
galaxies as those having $3\le$EW(H$\delta$)$\le$8 \AA $\,$ and
EW(H$\delta$)$>$8 \AA, respectively, and no emission lines (the
``post-starbust'' galaxies of Couch et al. \cite{cou94});
``e(a)''-type galaxies as those having emission lines and
EW(H$\delta$)$\ge$4 \AA$\,$; ``e(c)''-type galaxies as those having
emission lines and EW(H$\delta$)$<$4 \AA$\,$ (probably spiral
galaxies).  Galaxy ID 89, which is the only one showing any emission
lines ([OIII] in this case), but having the [OII] line outside the
observed spectral range, is classified as ``e(a)''.  Out of 106
classified galaxies (94 with $S/N\gtrsim10$) we find 68 ``passive''
galaxies and 38 ``active'' galaxies (starbust and poststarbust; 
i.e.\ 6:1:8:9:5:9 for k+a:a+k:e(c):e(a):e(b):e, respectively). Of 
nine galaxies of type ``e'', i.e.\ those moderate emission line 
galaxies with low $S/N$, six are probably of type e(c) and three 
are probably of type e(a) (hereafter e(c?) and e(a?)).

\subsection{Photometry}

Our photometric observations were carried out with the Wide Field
Camera (WFC), mounted at the prime focus of the 2.5 m INT (located at
Roque de los Muchachos Observatory, La Palma, Spain). We observed A773
in December 2004 in photometric conditions with a seeing of about
1.4\arcs.

The WFC consists of a four-CCD mosaic covering a
33\arcmm$\times$33\arcm field of view, with only a 20\% marginally
vignetted area. We took nine exposures of 720 s in $B_{\rm H}$ and 300
s in $R_{\rm H}$ Harris filters (a total of 6480 s and 2700 s in each
band) developing a dithering pattern of nine positions. This observing
mode allowed us to build a ``supersky'' frame that was used to correct
our images for fringing patterns (Gullixson \cite{gul92}). In
addition, the dithering helped us to clean cosmic rays and avoid gaps
between the CCDs in the final images. The complete reduction process
(including flat fielding, bias subtraction and bad-column elimination)
yielded a final coadded image where the variation of the sky was lower
than 0.4\% in the whole frame.  Another effect associated with the
wide field frames is the distortion of the field. In order to match
the photometry of several filters, a good astrometric solution is
needed to take into account these distortions. Using IRAF tasks and
taking as a reference the USNO B1.0 catalogue, we were able to find an
accurate astrometric solution (rms $\sim$ 0.5\arcss) across the full
frame. The photometric calibration was performed using Landolt
standard fields obtained during the observation.

We finally identified galaxies in our $B_{\rm H}$ and $R_{\rm H}$
images and measured their magnitudes with the SExtractor package
(Bertin \& Arnouts \cite{ber96}) and AUTOMAG procedure. In a few cases
(e.g.\ close companion galaxies, galaxies close to defects of the CCD)
the standard SExtractor photometric procedure failed. In these cases
we computed magnitudes by hand. This method consists in assuming a
galaxy profile of a typical elliptical and scaling it to the maximum
observed value. The integration of this profile gives us an estimate
of the magnitude. This method is similar to PSF photometry, but
assumes a galaxy profile, more appropriate in this case.

We transformed all magnitudes into the Johnson--Cousins system
(Johnson \& Morgan \cite{joh53}; Cousins \cite{cou76}). We used
$B=B\rm_H+0.13$ and $R=R\rm_H$ as derived from the Harris filter
characterization
(http://www.ast.cam.ac.uk/$\sim$wfcsur/technical/photom/colours/) and
assuming a $B-V\sim 1.0$ for E-type galaxies (Poggianti \cite{pog97}).
As a final step, we estimated and corrected the galactic extinction
$A_B \sim0.06$, $A_R \sim0.04$ from Burstein \& Heiles's
(\cite{bur82}) reddening maps.

We estimated that our photometric sample is complete down to $R=22.4$
(23.6) and $B=20.4$ (22.5) for $S/N=5$ (3) within the observed field.

We assigned magnitudes to 141  of the 142 galaxies of
our spectroscopic catalogue.  We measured redshifts for galaxies down to
magnitude $R\sim$ 21, but a high level of completeness is reached
only for galaxies with magnitude $R<$ 19.5 ($\sim$50\% completeness).

Table~\ref{tab1} (available in electronic format) lists the
velocity catalogue (see also Fig.~\ref{figimage} and
Fig.~\ref{figisofote}): identification number ID (Col.~1),
identification label of each galaxy following IAU nomenclature rules
(Col.~2); right ascension and declination, $\alpha$ and $\delta$
(J2000, Col.~3); $B$ magnitudes (Col.~4); $R$ magnitudes (Col.~5);
heliocentric radial velocities, ${\rm v}=cz_{\sun}$ (Col.~6) with
errors, $\Delta {\rm v}$ (Col.~7); spectral classification (Col.~8).

\section{Analysis of optical data}

\subsection{Member selection and global properties}

\begin{figure}
\centering
\resizebox{\hsize}{!}{\includegraphics{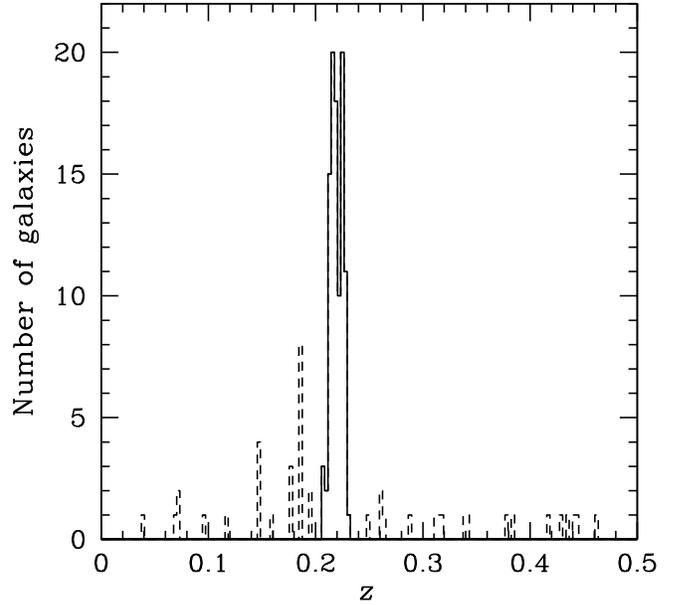}}
\caption
{Redshift galaxy distribution. The solid line histogram 
refers to galaxies assigned to the cluster according to
the DEDICA reconstruction method.}
\label{figden}
\end{figure}

To select cluster members from 142 galaxies having redshifts, we use
the adaptive-kernel method (hereafter DEDICA, Pisani \cite{pis93} and
\cite{pis96}; see also Fadda et al. \cite{fad96}; Girardi et
al. \cite{gir96}; Girardi \& Mezzetti \cite{gir01}). We find
significant peaks in the velocity distribution $>$99\% c.l..  This
procedure detects A773 as a two-peaked structure, populated by 100 
galaxies with $62250<{\rm v}<68850$ \kss, hereafter referred as cluster members 
(see Table~\ref{tab1} and Fig.~\ref{figden}).  Of the non-member galaxies,
24 and 18 are foreground and background galaxies, respectively. In
particular, 15 foreground galaxies belong to a low density peak at
$z\sim 0.186$.

By applying the biweight estimator to cluster members (Beers et
al. \cite{bee90}), we compute a mean cluster redshift of
$\left<z\right>=0.2197\pm$ 0.0005, i.e. $\left<\rm{\rm
v}\right>=65876\pm$140 \kss.  We estimate the LOS velocity dispersion,
$\sigma_{\rm v}$, by using the biweight estimator and applying the
cosmological correction and the standard correction for velocity
errors (Danese et al. \cite{dan80}).  We obtain $\sigma_{\rm
v}=1394_{-68}^{+84}$ \kss, where errors are estimated through a
bootstrap technique.

Hereafter, we consider as cluster centre the position of the most
luminous dominant galaxy [D1,
R.A.=$09^{\mathrm{h}}17^{\mathrm{m}}53\dotsec26$, Dec.=$+51\degree
43\arcmm 36\dotarcs5$ (J2000.0)], which is related to the richer
subsystem as shown in the following sections.

\subsection{Bimodal velocity structure}
\label{sec:velstr}

\begin{figure}
\centering 
\resizebox{\hsize}{!}{\includegraphics{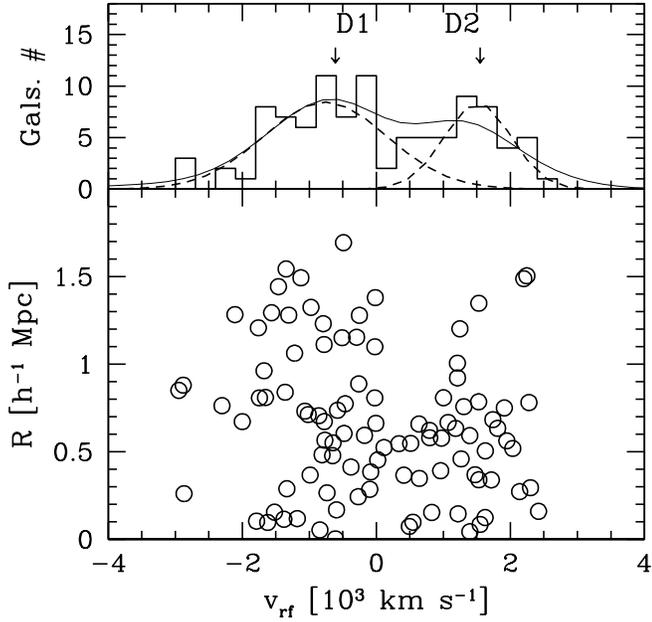}}
\caption
{ {\em Lower panel}: rest-frame velocity vs.\ projected distance from
the cluster centre -- the location of the dominant galaxy D1 -- of the 100
galaxies assigned to the cluster (Fig.~\ref{figden}).  {\em Upper
panel}: velocity histogram of the 100 galaxies assigned to the
cluster.  Velocities of the two dominant galaxies are pointed out.
The two--peaks detected by the DEDICA method and the two Gaussians
corresponding to the KMM partition (see Table~\ref{tabv}) are also
shown (faint and dashed lines, respectively).}
\label{figvd}
\end{figure}

The DEDICA method assigns 63 galaxies with ${\rm v}\leq 66543$ \ks to
a peak at $cz=65013$ \ks and 37 galaxies with ${\rm v}\geq 66653$ \ks
at a peak at $cz=67453$ ($z\sim 0.217$ and $z\sim 0.224$,
respectively, see Figure~\ref{figvd}).  According to the working
definition of Girardi et al. (\cite{gir96}), the two peaks are not
easily separable since their overlap is very large, i.e.\ many
galaxies (63/100) have a non-negligible probability of belonging
to both peaks.

The DEDICA method used above has the advantage of not requiring any a
priori shape for the subsystem research. Kaye's Mixture Model (KMM, as
implemented by Ashman et al. \cite{ash94}), which fits a
user-specified number of Gaussian distributions, gives a quite similar
result. Of the cluster members we find that the KMM method fits a
two-group partition by rejecting the single Gaussian at the 99.7\%
c.l., as obtained from the likelihood ratio test, assigning the 65
galaxies with ${\rm v}\leq 66659$ \ks and the 35 galaxies ${\rm v}\geq
66845$ \ks to two different groups.

Finally, we note that the two dominant galaxies of the cluster are
assigned to different groups: D1 lies very close to the low velocity
peak and D2 lies very close to the peak of the high velocity peak
(Fig.~\ref{figvd}).

We present the main properties for low- and high-velocity structures
as defined by both the DEDICA method (hereafter LV- and HV-groups,
respectively) and the KMM method (hereafter, KMM1 and KMM2,
respectively).  Table~\ref{tabv} lists the name of the sample
(Col.~1); the number of assigned members, $N_{\rm g}$ (Col.~2); the
mean velocity and its error, $<{\rm v}>$ (Col.~3); the velocity
dispersion and its bootstrap errors, $\sigma_{\rm v}$ (Col.~4).  The
above properties are computed using the galaxies assigned to each
individual group.  However, since there is a wide velocity range where
galaxies have a non-zero probability of belonging to both groups, both
DEDICA and KMM group membership assignment leads to an artificial
truncation of the distributions being ``deconvolved'' (see the
velocity thresholds above).  This truncation may lead to an
underestimate of velocity dispersion for the subclusters (see Bird
\cite{bir94}). Therefore, we also list the value of $\sigma_{\rm v}$
as computed by KMM software using the galaxies of the whole sample
weighted with their assignment probabilities to the respective sample
after having applied the cosmological correction and the standard
correction for velocity errors (Danese et al. \cite{dan80}).

\addtocounter{table}{+1}
\begin{table}
        \caption[]{Results of the kinematical analysis}
         \label{tabv}
                $$
         \begin{array}{l r l l}
            \hline
            \noalign{\smallskip}
            \hline
            \noalign{\smallskip}
\mathrm{Sample} & \mathrm{N_g} & 
\phantom{249}<{\rm v}>^{\mathrm{a}}\phantom{249} & 
\phantom{24}\sigma_{\rm v}^{\mathrm{b}}\phantom{24}\\
& &\phantom{249}\mathrm{km\ s^{-1}}\phantom{249} 
&\phantom{2}\mathrm{km\ s^{-1}}\phantom{24}\\
            \hline
            \noalign{\smallskip}
 
\mathrm{Whole\ system}&100 &65876\pm140 &1394_{-68}^{+84 }\\
\mathrm{LV-group}       & 63 &64875\pm105 & 825_{-63}^{+102}\\
\mathrm{HV-group}       & 37 &67664\pm86  & 515_{-43}^{+53 }\\
\mathrm{KMM1}         & 65 &64924\pm107 & 858_{-73}^{+96 } (1122)^{\mathrm{c}}\\
\mathrm{KMM2}         & 35 &67717\pm83  & 484_{-40}^{+50 } (548)^{\mathrm{c}}\\
\mathrm{W-sub}        & 14 &65573\pm366 &1299_{-118}^{+220}\\
\mathrm{E-sub}        &  5 &67469\pm86  & 473_{-25}^{+286 }\\
\mathrm{LV-W-sub}     &  8 &64415\pm198 & 493_{-43}^{+107}\\
\mathrm{HV-W-sub}     &  6 &67187\pm262 & 539_{-50}^{+84}\\
\mathrm{LV-very\ red}  & 28 &64962\pm100 & 518_{-64}^{+76}\\
\mathrm{HV-very\ red}  & 15 &67579\pm138 & 506_{-97}^{+149}\\
              \noalign{\smallskip}
            \hline
            \noalign{\smallskip}
            \hline
         \end{array}
$$
\begin{list}{}{}  
\item[$^{\mathrm{a}}$]For comparison $\rm v$ of D1 and D2 dominant
galaxies are 65131 and 67765 km$\ $s$^{-1}$.
\item[$^{\mathrm{b}}$] We use the biweight and the gapper estimators
by Beers et al. (\cite{bee90}) for samples with $\mathrm{N_g}\ge$ 15 and
with $\mathrm{N_g}<15$ galaxies, respectively (see also Girardi et
al. \cite{gir93}).
\item[$^{\mathrm{c}}$] Between parenthesis we list (standard) velocity
dispersions as computed by using the galaxies of the whole system
weighted with their assignment probabilities to the respective sample
rather then only the galaxies assigned to the samples (see text).
\end{list}
         \end{table}

Since the velocity difference between the two groups is probably
of a dynamical nature (see Sect.~5) we expect that the magnitude
incompleteness of our sample affects both in a similar way, thus
having a small influence on their detection and kinematic
properties. To verify this issue, of 142 galaxies having redshifts we
select samples of 98 (62) galaxies brighter than 19.5 (19) mags for
which the completeness is at least $\sim 50$\%. In both cases the KMM
method detects the two peaks at similar mean velocities. The estimates
of velocity dispersions agree within 1--1.5 sigma with the values
obtained in the whole sample (i.e.\ $\sigma_{\rm v}=750_{-85}^{+146}$,
$671_{-50}^{+110}$ \ks and $\sigma_{\rm v}=514_{-68}^{+82}$,
$474_{-118}^{+126}$ \ks for the main and secondary peaks, respectively
-- to be compared with KMM1 and KMM2 values).

\subsection{2D galaxy distribution}

\begin{figure}
\centering
\includegraphics[width=9cm]{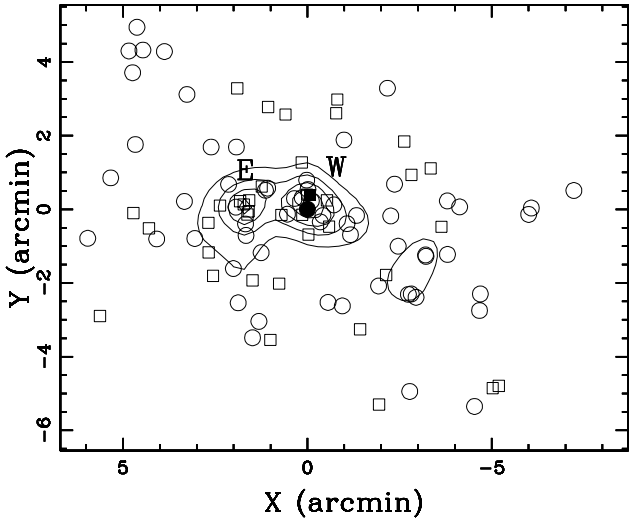}
\caption
{Spatial distribution on the sky of spectroscopically confirmed
cluster members and the relative isodensity contour map.  The contour
map shows two main peaks: the main, western peak (W-peak) and the
secondary, eastern peak (E-peak).  The cluster members are denoted by
circles (LV-galaxies) and squares (HV-galaxies) and the two dominant
galaxies are shown by solid symbols. The plot is centred on the
cluster centre defined as the position of the 
dominant galaxy D1 (solid circle).}
\label{figk2z}
\end{figure}

By applying the DEDICA method to the 2D distribution of A773 galaxy
members we find two peaks. The position of the highest peak (hereafter
W-peak) is very close to the location of the two dominant galaxies and
of the peak of X-ray emission (see Sect.~4), while the secondary peak
(hereafter E-peak) lies $\sim 2$\arcm east
[R.A.=$09^{\mathrm{h}}17^{\mathrm{m}}53\dotsec22$, Dec.=$+51\degree
43\arcmm 46\dotarcs0$ and
R.A.=$09^{\mathrm{h}}18^{\mathrm{m}}04\dotsec15$, Dec.=$+51\degree
43\arcmm 40\dotarcs9$ (J2000.0), respectively, see Fig.~\ref{figk2z}].

\begin{figure}
\centering
\includegraphics[width=8cm]{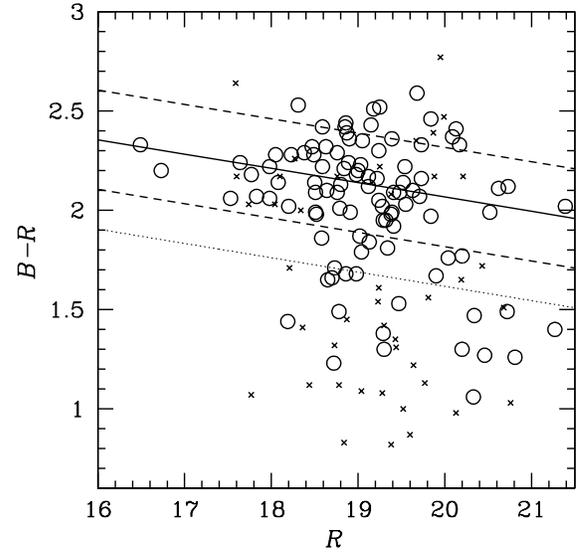}
\caption
{ $B-R$ vs.\ $R$ diagram for galaxies with available spectroscopy is
shown by circles and small crosses (cluster and field members,
respectively).  The solid line gives the best-fit colour--magnitude
relation as determined from member galaxies; the dashed lines are
drawn at $\pm$0.25 mag from the CMR. According to our working
definition in Sect.~3.5, cluster members are divided into ``very
red'', ``red'' and ``blue'' galaxies (above the CMR, between the CMR
and the dotted line and below the dotted line, respectively).  }
\label{figcm}
\end{figure}

\begin{figure}
\centering
\includegraphics[width=8cm]{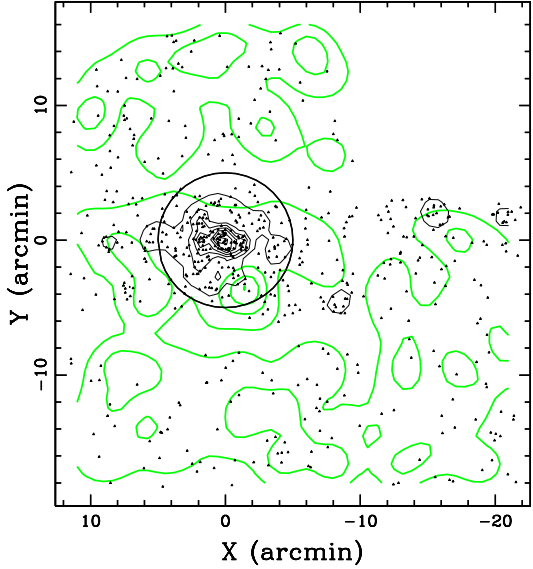}
\caption
{Spatial distribution on the sky and relative isodensity contour map
of 582 likely cluster members (according to the colour--magnitude
relation) with $R\le 21$, obtained with the DEDICA method (black
lines).  For comparison we show the contour map of the 563 likely
non-cluster members (grey lines).  The plot is centred on the cluster
centre.  The circle indicates the 5\arcm central region. }
\label{figk2A}
\end{figure}

Our spectroscopic data do not cover the entire cluster field and
suffer from magnitude incompleteness. To overcome these limitations we
recover our photometric catalogue by selecting likely members on the
basis of the colour--magnitude relation (hereafter CMR), which
indicates the early-type galaxy locus.  To determine the CMR we fix
the slope according to L\'opez--Cruz et al. (\cite{lop04}, see their
fig.~3) and apply the two-sigma-clipping fitting procedure to the
cluster members obtaining $B-R=3.507-0.072\times R$ (see
Fig.~\ref{figcm}).  From our photometric catalogue we consider
galaxies (objects with SExtractor stellar index $\le 0.9$) lying
within 0.25 mag of the CMR.  To avoid contamination by field galaxies
we do not show results for galaxies fainter than 21 mag (in the
$R$-band).  The contour map for 582 likely cluster members having
$R\le 21$ shows how the main structure is elongated in the EW
direction with two main peaks corresponding to those determined above
(see Fig.~\ref{figk2A}).  To check the effect of the gaps between the
CCD chips of the WFC we also consider the distribution of 563 galaxies
having $R\le 21$ and lying more than 0.75 mag from the
colour--magnitude relation, i.e.\ probably mainly formed by non-member
galaxies. The comparison of members and non-members in
Fig.~\ref{figk2A} confirms the reality of the above two peaks in the
cluster structure.

\begin{figure}
\centering
\includegraphics[width=8cm]{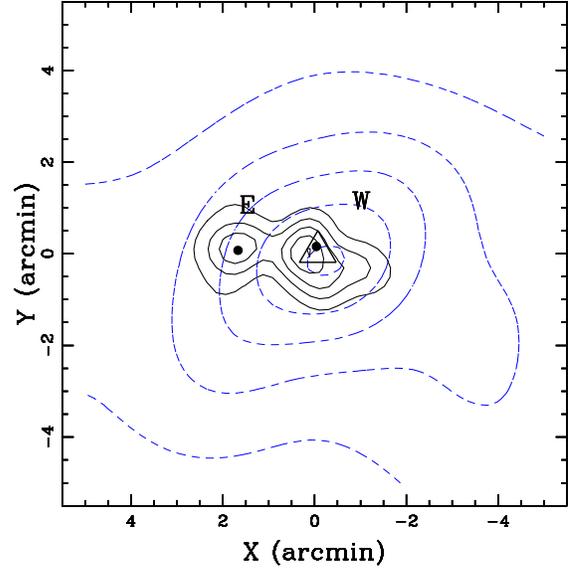}
\caption
{Isodensity contour map (zoomed within 1 \h from the cluster centre)
of likely cluster members of different magnitude intervals. For the
sake of clarity we only show those for $R\le 19$ and $19<R\le 20$,
indicated by black solid lines and dashed lines.  We also indicate the
position of the X-ray centre (large triangle), and the E- and W-peaks
as recovered from galaxies having redshifts (small dots, see also
Fig.~\ref{figk2z}).}
\label{figk2B}
\end{figure}

When analysing the 2D galaxy distribution for magnitude intervals the
bimodal structure in the central cluster region is confirmed by the
141 galaxies brighter than 19 mag (in the $R$-band), while two samples of
fainter galaxies (177 with $19<R\le20$ and 262 with $20<R\le21$) show
only one significant peak closer to the W-peak than to the E-peak 
(see Figure~\ref{figk2B}).  

\subsection{3D structure of A773}
\label{sec:3dstr}

The existence of correlations between positions and velocities of
cluster galaxies is a footprint of real substructures.

In the case of A773 the standard approaches to determine these
correlations fail (e.g.\ Boschin et al.~\cite{bos06}; Girardi et
al.~\cite{gir06}). In fact, we find no significant velocity gradient,
no significant substructure according to the method by Dressler \&
Schectman (\cite{dre88}), and the result of the 3D KMM method is fully
driven by the velocity-variable making us obtain again the two-group
partition of Section~3.2.

As an alternative approach, we analyse the 2D distributions of the two
groups determined using the velocity distribution.  We compare the 2D
distribution of the galaxies assigned to the LV-group (hereafter
LV-galaxies) with that of the galaxies assigned to the HV-group
(hereafter HV--galaxies) finding no significant difference according
to the 2D Kolmogorov--Smirnov tests (2DKS test, see Fasano \&
Franceschini \cite{fas87} as implemented by Press et
al. \cite{pre92}).

\begin{figure}
\centering
\includegraphics[width=8cm]{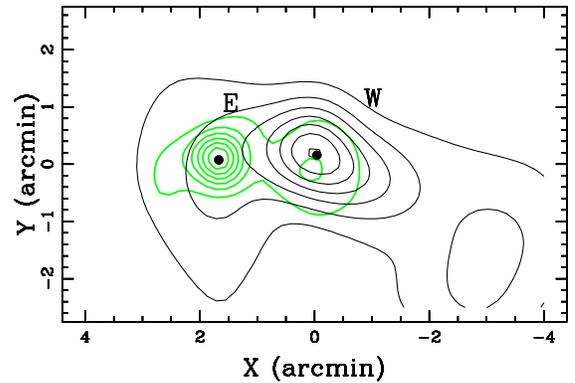}
\caption
{Comparison of isodensity contour maps recovered from the galaxies
assigned to the low velocity peak (black solid line) and from galaxies
assigned to the high velocity peak (light grey line).  The position of
peaks W and E are indicated by solid circles.}
\label{figk2p}
\end{figure}

To go deeper into the question we also use the 2D DEDICA method. We
find that the distribution of LV-galaxies shows a main, very
significant peak located close to the W-peak, while the main 2D peak
of the distribution of HV-galaxies is closely located near the E-peak,
thus indicating a relation between velocity structure and 2D structure
detected in the above two sections (see Fig.~\ref{figk2p}).

Moreover, the distribution of HV-galaxies also shows a secondary peak
closely located near the W-peak. This secondary peak is only 93\%
significant and much less dense with respect to the main one (only
30\%), but contains a similar number of galaxies (seven vs.\ eight).
This secondary peak is also found when we consider only the 22
galaxies with a velocity higher than the peak velocity of $67453$ \ks
and therefore we exclude the possibility that the 2D secondary peak is
induced by the difficulties in member assignment between the two
velocity peaks (see Sect.~3.2).  The likely conclusion is that some
HV-galaxies are really connected with the high velocity D2 galaxy
located in the region of the W-peak.

As a final approach, we analyse the velocity distributions of the
galaxy populations located in the regions of the E- and W-peaks.
Following Girardi et al. (\cite{gir05}) we analyse the profiles of
mean velocity and velocity dispersion of galaxy systems corresponding
to the E- and W-peaks to attempt an independent analysis.

\begin{figure}
\centering 
\resizebox{\hsize}{!}{\includegraphics{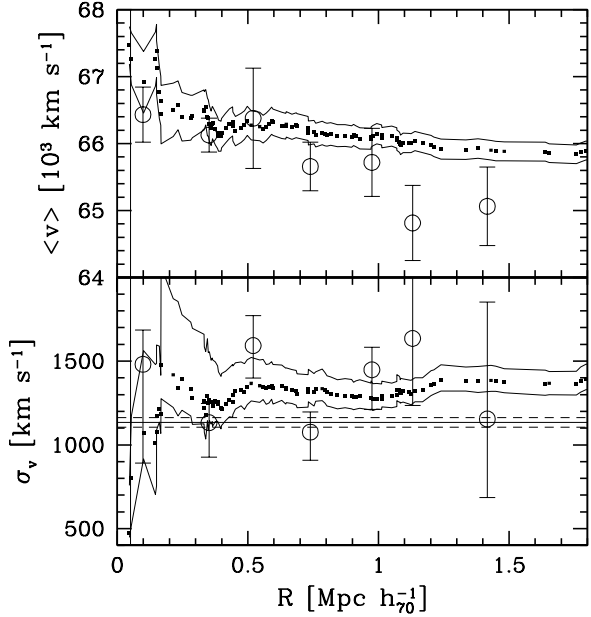}}
\caption
{Kinematic profiles of the cluster obtained assuming the E-peak as
centre. The vertical line indicates the region which encloses the
galaxies of E-sub (see text).  Differential (big circles) and integral
(small points) mean velocity and LOS velocity dispersion profiles are
shown in top and bottom panels, respectively. For the differential
profiles are plotted the values obtained binning galaxies 10 by
10. For the integral profiles, the mean and dispersion at a given
(projected) radius from the position of the E-peak is estimated by
considering all galaxies within that radius: the first point refers to
the first five galaxies and the last point refer to all galaxies
contained in the cluster.  The error bands at the $68\%$ c.l. are also
shown.}
\label{figprofc2}
\end{figure}

\begin{figure}
\centering 
\resizebox{\hsize}{!}{\includegraphics{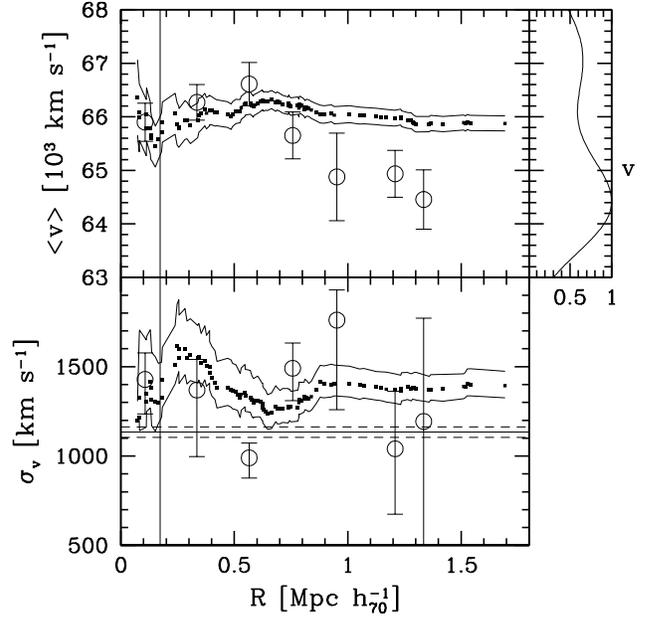}}
\caption
{The same as in Fig.~\ref{figprofc2}, but referring to the W-peak.
The small upper-right panel shows the velocity galaxy density (in
normalized units), as provided by DEDICA for the galaxies of W-sub,
i.e.\ those galaxies contained within the radius indicated by the
vertical line.}
\label{figprofc1}
\end{figure}

In the case of the E-peak the integral $\sigma_{\rm v}$ profile shows a
sharp increase starting from the peak centre (see
Fig.~\ref{figprofc2}).  This is probably due to the contamination of a
galaxy clump with a different mean velocity (see also Girardi et
al. \cite{gir96} in the case of A3391--A3395). To avoid possible
problems of contamination we decide to consider a sample E-sub formed
by only the first five galaxies closest to the E-peak.  In the case
of the W-peak the value of $\sigma_{\rm v}$ computed using the first five
galaxies is enough high -- $\sim 1200$ \ks -- and the integral
$\sigma_{\rm v}$ profile shows an almost stable behaviour out to 
$\sim$0.2 \h , where it starts to increase (see Fig.~\ref{figprofc1}).  We
decided to consider a sample W-sub formed by the first fourteen
galaxies closest to the W-peak.

Table~\ref{tabv} lists kinematic properties for E- and W-subs and
shows that these groups have similar properties of the 
HV-subcluster and of the whole cluster, respectively. When comparing 
the kinematics of E-sub and W-sub we find that their velocity
distributions are different at the 97.8\% c.l. according to the KS
test (e.g.\ Ledermann \cite{led82}, as implemented by Press et
al. \cite{pre92} for small samples), and that the mean velocity and
the velocity dispersion differ at the 99.6\% c.l and 93.5\% c.l.
according to the means test and F test, respectively (e.g.\ Press et
al. \cite{pre92}).  Moreover, W-sub has a complex structure since its
velocity distribution shows two peaks according to DEDICA (see the
small upper-right panel of Fig.~\ref{figprofc1}).  The low velocity
peak (LV-W-sub) contains the D1 galaxy and resembles the mean velocity
of LV-subcluster.  The high velocity peak (HV-W-sub) contains the D2
galaxy and resembles the mean velocity of HV-subcluster.

\subsection{Further insights using galaxy populations}


In the case of A773, strongly substructured, the analysis of red
and/or bright galaxies might be useful to trace the cores of the
subclusters (e.g.\ Lubin et al. \cite{lub00}; Boschin et
al. \cite{bos04}). Moreover, active star forming galaxies might be
related to the cluster--cluster merging phenomena (e.g.\ Bekki
\cite{bek99}).

\begin{figure}
\centering
\resizebox{\hsize}{!}{\includegraphics{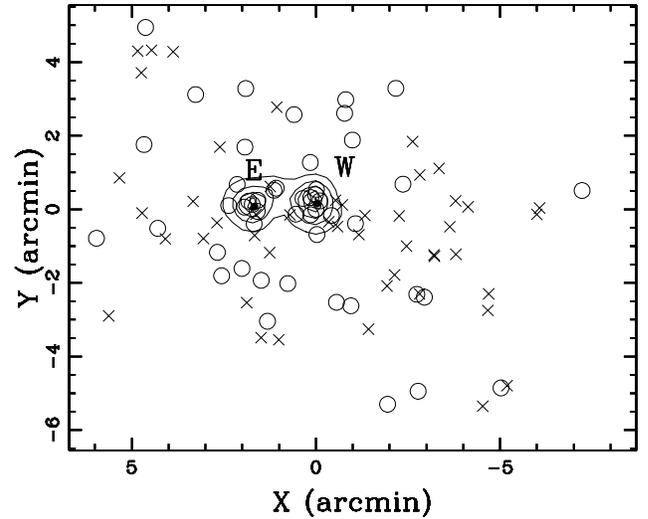}}
\caption
{Spatial distribution on the sky of 100 cluster members.  More and
less luminous galaxies are denoted by circles and crosses,
respectively.  The contour map recovered from the most luminous
galaxies is also shown.  The position of peaks W and E are indicated
by solid, small circles.}
\label{figk2l}
\end{figure}

We divide the sample (99 galaxies having magnitudes) into low- and
high-luminosity subsamples, each with 49 galaxies, by using the median
$R$ magnitude = 19.08.  The two subsamples do not differ in their
velocity distribution according to the KS test and both show a bimodal
velocity distribution according to the DEDICA method.  In agreement
with the photometric sample results, luminous galaxies clearly show
the presence of both E and W peaks (see Figure~\ref{figk2l}).

\begin{figure}
\centering
\resizebox{\hsize}{!}{\includegraphics{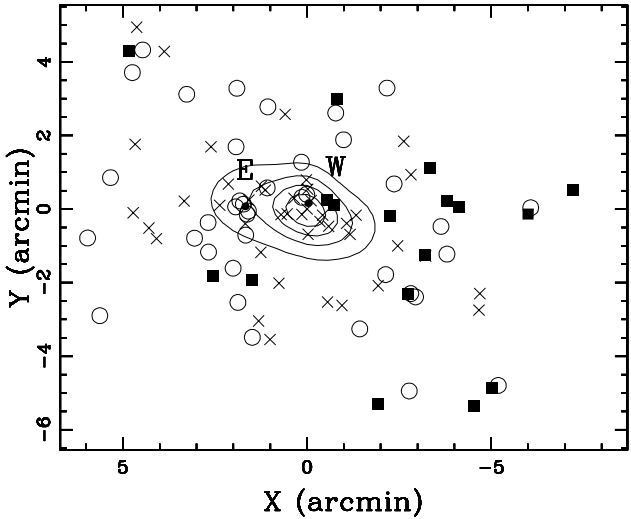}}
\caption
{Spatial distribution on the sky of 100 cluster members.  Very red,
red and and blue galaxies are denoted by crosses, circles and solid
squares, respectively.  The contour map recovered from the very red
galaxies is also shown.  The position of peaks W and E are indicated
by solid small circles.}
\label{figk2red}
\end{figure}

Butcher \& Oemler (\cite{but84}) define as ``blue'' those galaxies
with rest-frame $B-V$ colours at least 0.2 mag bluer than that of the
CMR, this translates to 0.447 mag for $B-R$ colours at $z\sim 0.21$
(see Mercurio et al. \cite{mer04} and Haines et al. \cite{hai04} for
A209), and a comparable result is obtained by using the tables of
Fukugita et al.  (\cite{fuk95}).  Therefore, we hereafter define
$(B-R)_{\rm corr}=(B-R)-(3.507-0.072\times R)$ and ``blue'' those
galaxies with $(B-R)_{\rm corr}<-0.45$, i.e.\ those with observed
$B-V$ colours at least 0.45 mag bluer than that of the CMR.  To
analyse the colour segregation we also divide other galaxies into
``red'' [$-0.45\le(B-R)_{\rm corr}\le0$] and ``very red'' [$(B-R)_{\rm
corr}>0$]. We obtain three subsamples of 43 very red, 39 red, and 17
blue galaxies (see Fig.~\ref{figcm}).  The subsample of the very red
galaxies shows two distinct peaks in the velocity distribution
(LV--very red and HV--very red, see Table~\ref{tabv}).  When analysing
the 2D galaxy distribution of each subsample through the DEDICA method
we find that the very red galaxies show a significant peak at the
position of the W-peak -- this is true for both LV--very red and
HV--very red subsamples separately -- while red galaxies show a
significant peak at the position of the E-peak (see
Fig.~\ref{figk2red}).  The blue galaxies are instead preferentially
located towards western regions of the cluster.

\begin{figure}
\centering
\resizebox{\hsize}{!}{\includegraphics{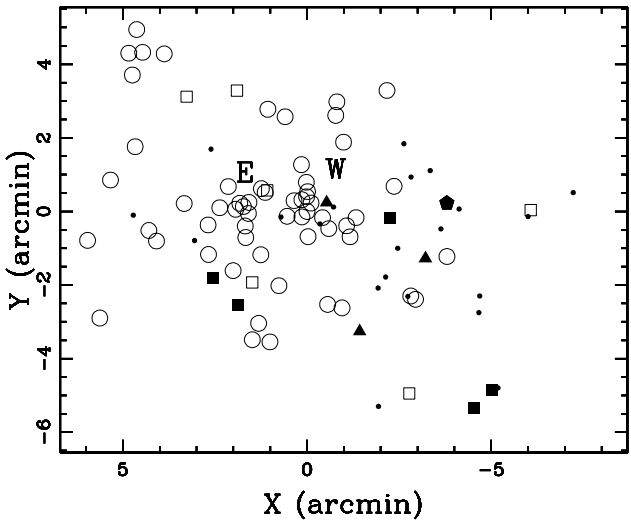}}
\caption
{Spatial distribution on the sky of 100 cluster members.  Large
symbols denote spectroscopically classified galaxies: circles,
squares, solid triangles, solid squares and solid pentagons indicate
k,k+a/a+k,e(c)/e(c?),e(a)/e(a?) and e(b), respectively.  Small dots
indicate unclassified members.}
\label{figk2spec}
\end{figure}

Finally, we consider galaxies of different spectral types.  Of 76
classified member galaxies, we find 60 ``passive'' k-galaxies and 16
``active'' galaxies, of which 10 are emission line galaxies
(i.e.\ 5:1:1:3:1:5 for k+a:a+k:e(c)/e(c?):e(a)/e(a?):e(b):e,
respectively). Both passive and emission line galaxy populations show
the characteristic bimodal velocity distribution and have similar mean
velocity and velocity dispersion, quite comparable to that of the
whole sample.  As for the 2D distribution, the passive population
shows the characteristic bimodal distribution with the E- and W-peaks,
while the active populations avoid these regions of high density.  In
particular, Fig.~\ref{figk2spec} shows that emission line galaxies are
preferentially located towards south-western regions of the cluster.
Indeed, we find that passive galaxies differ from the emission-line
galaxies at the 98.0\% c.l., respectively, according to the 2DKS test
and despite some categories drawn in Fig.~\ref{figk2spec} include 
small numbers of objects and make this comparison difficult.

\section{X-ray analysis}
\label{sec:xray}

X-ray analysis of A773 has already been performed in the past, both
based on ROSAT (Pierre \& Starck \cite{pie98}, Rizza et
al. \cite{riz98}, Govoni et al. \cite{gov01b}) and Chandra data
(Govoni et al. \cite{gov04}). In particular, Govoni et
al. (\cite{gov04}) analysed two Chandra ACIS-I observations (exposures
ID \#533 and \#3588). The two pointings have a clean total exposure
time of 18.7 ks (see their table 1). More recently, Chandra has again
observed A773 (exposure ID \#5006). This pointing has an exposure time
of 20.1 ks. In this section we propose an analysis of this Chandra
ACIS-I observation.

Data reduction is performed using the package CIAO\footnote{CIAO is
freely available at http://asc.harvard.edu/ciao/} (Chandra Interactive
Analysis of Observations) on chip I3 (field of view $\sim
8.5\arcmin\times 8.5\arcmin$). First, we remove events from the level
2 event list with a status not equal to zero and with grades one, five
and seven. We then select all events with energy between 0.3 and 10
keV. In addition, we clean bad offsets and examine the data, filtering
out bad columns and removing times when the count rate exceeds three
standard deviations from the mean count rate per 3.3 s interval. We
then clean the I3 chip for flickering pixels, i.e.\ times where a
pixel has events in two sequential 3.3 s intervals. The resulting
exposure time for the reduced data is 19.6 ks.

In Fig.~\ref{figisofote} we plot an $R$-band image of the cluster with
superimposed the X-ray contour levels of the Chandra image. The shape
of the cluster appears to be moderately elliptical. By using the CIAO
package Sherpa we fit an elliptical 2D Beta model to the X-ray photon
distribution to quantify the departure from the spherical shape. The
model is defined as follows: 
\begin{equation}
f(x,y)=f(r)=A/[1+(r/r_0)^2]^{\alpha}, 
\end{equation} 
where the radial coordinate $r$ is defined as
$r(x,y)=[X^2(1-\epsilon)^2+Y^2]^{1/2}/(1-\epsilon)$,
$X=(x-x_0)\,\cos\,\theta+(y-y_0)\,\sin\,\theta$ and
$Y=(y-y_0)\,\cos\,\theta-(x-x_0)\,\sin\,\theta$. Here $x$ and $y$ are
physical pixel coordinates on chip I3. The best fit centroid position
is located very close to the dominant galaxy D1,
i.e. R.A.=$09^{\mathrm{h}}17^{\mathrm{m}}52\dotsec87$,
Dec.=$+51\degree 43\arcmm 39\dotarcs2$ (J2000.0).  The best-fit core
radius, the ellipticity and the position angle are $r_0=47\pm$3 arcsec
(167$\pm$11 \kpcc), $\epsilon=0.23\pm$0.02 and PA=79.7$\pm$2.8 degrees
(measured counterclockwise from the north), respectively.

Isophotes in Fig.~\ref{figisofote} suggest that the core of A773 is
substructured with two distinct maxima in the surface brightness
distribution. The presence of these two peaks is confirmed by a
wavelet analysis of the image performed by the task
CIAO/Wavdetect. The task was run on several scales to search for
substructures with different sizes. The significance
threshold\footnote{see \S~11.1 of the CIAO Detect Manual (software
release version 3.3, available at the WWW site
http://cxc.harvard.edu/ciao/manuals.html)} was set at $10^{-6}$.  The
results are shown in Fig.~\ref{figisofote}. Green ellipses represent
two significant surface brightness peaks found by Wavdetect in the
core of the cluster. They are located at
R.A. $09^{\mathrm{h}}\,17^{\mathrm{m}}\,53.4^{\mathrm{s}}$ and
Dec. $+51\degr\,43\arcmin\,44\arcsec$, and at
R.A. $09^{\mathrm{h}}\,17^{\mathrm{m}}\,51.4^{\mathrm{s}}$ and
Dec. $+51\degr\,43\arcmin\,32\arcsec$, respectively.  This finding
confirms the results of Pierre \& Stark (\cite{pie98}) and Rizza et
al. (\cite{riz98}), who first noted the complex structure of the core
from a 16 ks ROSAT/HRI exposure.

For the spectral analysis of the cluster X-ray photons, we compute a
global estimate of the ICM temperature. The temperature is computed
from the spectrum of the cluster within a circular aperture of
2$\arcmin$ radius around the cluster centre. Fixing the absorbing
galactic hydrogen column density at 1.44$\times$10$^{20}$ cm$^{-2}$,
computed from the HI maps by Dickey \& Lockman (\cite{dic90}), we fit
a Raymond--Smith (\cite{ray77}) spectrum using the CIAO package Sherpa
with a $\chi^{2}$ statistics. Considering the energy range 0.8--10
keV, we find a best fit temperature of $T_{\rm X}=\,$7.8\,$\pm\,0.4$
keV and a metal abundance of 0.34$\pm\,0.07$ solar, in agreement 
with the results of Govoni et al. (\cite{gov04}).

\section{Discussion}



The global LOS velocity dispersion $\sigma_{\rm v}=1394_{-68}^{+84}$
\ks is considerably higher than the average X-ray temperature $T_{\rm
X}=$7.8$\pm$0.4 keV from our analysis of Chandra data when assuming
the equipartition of energy density between gas and galaxies (i.e.\
$\beta_{\rm spec} =1.51^{+0.13}_{-0.10}$ to be compared with
$\beta_{\rm spec}=1$\footnote{$\beta_{\rm spec}=\sigma_{\rm
v}^2/(kT/\mu m_{\rm p})$ with $\mu=0.58$ the mean molecular weight and
$m_{\rm p}$ the proton mass.}). This suggests that the cluster
is far from dynamical equilibrium and indeed we do find evidence that
A773 shows two peaks both in the velocity space ($\Delta {\rm v} \sim
2800$ \kss) and in the 2D projected position space ($\sim 2\arcmin\,$
between the E-peak and the W-peak).

The main subsystem is detected as the LV-peak in the velocity
distribution, has a velocity dispersion of $\sigma_{{\rm v},{\rm
c}}$=800--1100 \ks\ -- comparable to massive clusters - and its centre
probably coincides with the D1 galaxy, located close to the main 2D
peak (W-peak) and the peak of X-ray emission. The distribution of very
red galaxies is also peaked in the same place.  Both the low velocity
galaxies located close to the W-peak (LV-W-sub) and the low velocity,
very red galaxies (LV-very red) show a $\sigma_{\rm v}$ value of about
500 \kss, lower than the global sigma as often observed in the case of
a core cluster, in particular when examining passive galaxies (Adami
et al. \cite{ada98}; Biviano \& Katgert \cite{biv04}).

A secondary subsystem is detected as the HV-peak in the velocity
distribution and has a moderate velocity dispersion of $\sigma_{{\rm
v},{\rm g}}\sim500$ \kss. Its spatial structure is rather complex,
showing two galaxy concentrations, one at the location of the E-peak
and the other at the location of the W-peak, i.e.\ close to the high
velocity, D2 dominant galaxy (see Fig.~\ref{figk2p} and the
upper-right panel of Fig.~\ref{figprofc1}).  The absence of a
concentration of very red galaxies and of any close X-ray peak, as
well as the presence of only a minor DM peak (Dahle et
al. \cite{dah02}), suggest that the E-peak is dynamically of minor
importance. Analysing galaxies around the two concentrations (E-sub
and HV-W-sub) we find similar kinematic properties; thus, with present
data we cannot distinguish whether they are two independent galaxy
groups or two components of the same group.  In favour of the first
hypothesis, we notice that both theoretical predictions and
observational data suggest that clusters are formed by numerous small
infalling systems (see the case of A85-A87 complex in Durret et
al. \cite{dur98}). In this case, the similarity of the mean velocities
of E-sub and HV-W-sub might reflect that their velocities are of
dynamical origin, both groups strongly interacting with the same
cluster potential and thus having a similar merging velocity. In the
case of the second hypothesis, E-sub could be either a dense
substructure of the HV-system or a part of the core of the HV-system
dynamically decoupled from the dominant galaxy during the cluster
merger. We notice that Adami et al. (\cite{ada05}) found a similar
feature in the Coma cluster, where the N03--SW group has the same mean
velocity as NGC 4889, one of the two dominant galaxies, even if it is
spatially distant in 2D.

\subsection{A773 mass estimate}
\label{sec:mass}

On the basis of the above discussion we estimate the mass of A773,
assuming that the system is formed by a main cluster with a velocity
dispersion of $\sigma_{{\rm v},{\rm c}}$=800--1100 \ks and one or two
groups with $\sigma_{{\rm v},{\rm g}}\sim500$ \kss. Here we report our
results in the case of one group, but notice that in the case of two
groups, they have the same virial masses and the results of this
section are still valid.

Although the system is probably in a phase of strong interaction (see
also the following section) so that the subsystems are still
detectable in the velocity distribution, let us assume having a good
estimate of $\sigma_{\rm v}$ of the subsystems before the merger. To
apply the virial method we must assume that each subsystem is in
dynamical equilibrium.  This might be not true in the case of a
possibly substructured group (see above).  On the other hand, the mass
estimate of the whole complex depends above all on the main
subsystem. whose structure is much more regular.

Following the prescriptions of Girardi \& Mezzetti (\cite{gir01}), we
assume for the radius of the quasi-virialized region R$_{\rm
vir,c}=0.17\times \sigma_{\rm v}/H(z) = 2.3$-3.2 \h and R$_{\rm
vir,g}=0.17\times \sigma_{\rm v}/H(z) \sim 1.4$ \h for the cluster and
the group, respectively -- see their eq.~1 after introducing the
scaling with $H(z)$ (see also eq.~ 8 of Carlberg et al. \cite{car97}
for R$_{200}$). Therefore, our spectroscopic catalogue samples the
whole group virialized region and more than half of the cluster one.

One can compute the mass using the virial theorem (Limber \& Mathews
\cite{lim60}; see also, for example, Girardi et al. \cite{gir98})
under the assumption that mass follows galaxy distribution: $M=M_{\rm
svir}-\rm{SPT}$, where $M_{\rm svir}=3\pi/2 \cdot \sigma_{\rm v}^2{\rm
R}_{\rm PV}/G$ is the standard virial mass, R$_{\rm PV}$ a projected
radius (equal two times the harmonic radius), and SPT is the surface
pressure term correction (The \& White \cite{the86}).  The estimate of
$\sigma_{\rm v}$ is generally robust when computed within a large
cluster region (see Fig.~5 of Girardi et al. \cite{gir06} and Fadda et
al. \cite{fad96} for other examples), hence for each of the two
subsystems we consider the corresponding above-global values
$\sigma_{{\rm v},{\rm c}}$ and $\sigma_{{\rm v},{\rm g}}$.  The value
of R$_{\rm PV}$ depends on the size of the region considered so that
the computed mass increases (but not linearly) with the increasing
region considered.  Since the two subsystems are partially aligned
along the LOS we avoid computing R$_{\rm PV}$ by using data for
observed galaxies and use an alternative estimate that was shown to be
good when R$_{\rm PV}$ is computed within R$_{\rm vir}$ (see eq.~13 of
Girardi et al. \cite{gir98}).  This alternative estimate is based on
our knowledge of the galaxy distribution and, in particular, a galaxy
King-like distribution with parameters typical of
nearby/medium-redshift clusters: a core radius R$_{\rm
core}=1/20\times {\rm R}_{\rm vir}$ and a slope-parameter $\beta_{\rm
fit}=0.8$, i.e.\ the volume galaxy density at large radii as $r^{-3
\beta_{\rm fit}}=r^{-2.4}$ (Girardi \& Mezzetti \cite{gir01}).  For
the whole virialized region we obtain R$_{\rm PV,c}=1.7$-2.4 \h and
R$_{\rm PV,g}\sim1.4$ \h, where a 25\% error is expected because
typical, rather than individual, galaxy distribution parameters are
assumed. As for the SPT correction, we assume a 20\% one, computed
combining data on many clusters (e.g.\ Carlberg et al. \cite{car97};
Girardi et al. \cite{gir98}).  This leads to virial masses for the
subsystems of $M_{\rm c}(<{\rm R}_{\rm vir,c}=2.3-3.2\,\hhh)=1.0$-2.5
\mqui of $M_{\rm g}(<{\rm R}_{\rm vir,g}=1.4\,\hhh)\sim2.4$ \mqua with
a typical error of $\sim$30\%.  A reliable mass estimate of the whole
system is then $M=1.2$--2.7 \mquii, in agreement with rich clusters
reported in the literature (e.g.\ Girardi et al. \cite{gir98}; Girardi
\& Mezzetti \cite{gir01}).

To compare our result with the estimate obtained via gravitational
lensing we obtain a projected mass assuming that the cluster is
described by a King-like mass distribution (see above), or a NFW
profile where the mass-dependent concentration parameter is taken from
Navarro et al. (\cite{nav97}) and rescaled by the factor $1+z$
(Bullock et al. \cite{bul01}; Dolag et al. \cite{dol04}). Moreover, we
assume that the two subsystems have coincident centres, and that the
subsystems themselves are extended only to one R$_{\rm vir}$.  We
obtain $M_{\rm proj}(<\rm{R}=90$\arcss)=(4.4--7.9)\mqua in agreement
with that found by Dahle et al. (\cite{dah02}) considering the quoted
errors ($M_{\rm Dahle}\sim 3$\mqua with an upper error of the 50\%
after rescaling to our cosmological model).

Using the same mass distributions we compute $M(<\rm{R}=1$
\hh)=(5.9--11.1)\mqua and, using $r$-band luminosity by Popesso et
al.~(\cite{pop04}), estimate $M/L$=(150--280) \mll.  The upper value
is typical of clusters of similar redshift (Carlberg et
al.~\cite{car96}).

\subsection{Merging scenario}

The X-ray emission shows two very close peaks along the NE--SW
direction separated by $\sim$0.5\arcmm.  The main X-ray peak, the NE
one, is closely located near -- but is not coincident with -- the D1
galaxy.  The secondary, SW, one does not coincide with any luminous
galaxy or obvious galaxy substructure (see Fig.~\ref{figisofote}).
This lack of correlation between collisional and non-collisional
components strongly suggests that the cluster is undergoing an
advanced phase of merging (e.g.\ Roettiger et al. \cite{roe97}).
Moreover, the direction of elongation of the isophotes of the X-ray
emission (NEE--SWW direction, PA$\sim$80 degrees) is similar to that
indicated by the elongation of galaxy distribution (see, for example,
contours for very red galaxies in Fig.~\ref{figk2red}), but quite
different from N--S one indicated by the two dominant galaxies and the
elongation of the main DM clump (Dahle et al. \cite{dah02}). It is
worth noting that the literature reports several cases of clusters
having two dominant galaxies with significantly different velocities,
where the elongation of the X-ray emission is rotated from the axis
connecting the two dominants and/or with X-ray peaks not coinciding
with the location of the dominants (e.g.\ A2255 by Burns et
al. \cite{bur94}; RXJ1314-25 by Valtchanov et al.  \cite{val02}; A2744
by Boschin et al. \cite{bos06}).  On smaller scales, A773 also shows
minor but intriguing features.  The E-group is not really aligned with
the direction of the elongation of the X-ray emission, but rather
shows a PA $\sim$ 90 degrees with respect to the cluster
centre. Moreover, although the main DM peak roughly coincides with
that indicated by the two dominant galaxies, it indeed is offset by
about 1\arcmm.

Further support in favour of an advanced phase of merging comes from
the luminosity segregations. Although both luminous and faint galaxies
show two peaks in their velocity distribution, only the luminous ones
show the E-peak in the 2D distribution.  Indeed, it is not unusual
that galaxies of different luminosity trace the dynamics of cluster
mergers in a different way. The first evidence was given by Biviano et
al. (\cite{biv96}; see also Mercurio et al. \cite{mer03}), who found
that the two central dominant galaxies of the Coma cluster are
surrounded by luminous galaxies accompanied by the two main X-ray
peaks, while the distribution of faint galaxies tend to form a
structure not centred on one of the two dominant galaxies but rather
coincident with a secondary peak detected in X-rays.  Therefore,
following Biviano et al. (\cite{biv96}), we might speculate that the
merging is in an advanced phase, where faint galaxies trace the
forming structure of the cluster, while more luminous galaxies still
trace the remnants of pre-merging clumps, which could be so dense to
survive for a long time after the merging (as suggested by numerical
simulations; see Gonz\'alez--Casado et al. \cite{gon94}).

As discussed in the above sections, the subsystems involved in the
merger are one cluster and one or two groups with a large mass ratio
from 4:1 to 10:1, depending on which $\sigma_{\rm v}$ estimate we use
for the cluster, with a velocity separation that ranges $2100-2500$ 
\ks in the cluster rest-frame. 
Cosmological simulations suggest that these high impact velocities 
might be occurring in phenomena of cluster merging (e.g.\ Crone \& Geller
\cite{cro95}).  The value of a merging velocity of $\sim 3000$ \ks is
also predicted by $N$-body numerical simulations of Pinkney et
al. (\cite{pin96}) and is probably associated with the phase of the
core crossing in a cluster--cluster merging process.  In the case of
A773, the extremely large LOS component of the velocity indicates that
the LOS is close to the merging axis.

As for the direction of the merger, the elongations noted in the X-ray
emission, the galaxy and DM distributions indicate that the merger is
not occurring entirely along the LOS, but must have at least a small
transverse component in the plane of the sky. However, since the
variety of features do not indicate a unique direction we conclude
that the geometry of the merger is very complex.  The likely scenario
is that we are looking at a multiple merger where the N--S direction
might indicate the direction of an older merger event concerning the
cluster and the group associated with the D2 dominant galaxy. D1 and
D2 might be the tracers of the two cores destined to oscillate around
the mean velocity for a long time (i.e.\ from two to several $10^9$ yr
as suggested by numerical simulations; see, for example, Nakamura et
al. \cite{nak95}; Faltenbacher et al. \cite{fal06}). In fact, a high
velocity difference between the two dominant galaxies is often
suggestive of an energetic cluster merger (e.g.\ Burns et
al. \cite{bur95}) and this process is thought to be the cause of the
formation of dumbbell galaxies in a few merging clusters (e.g.\ Beers
et al. \cite{bee92}; Flores et al. \cite{flo00}). The NE by E--SW by W
elongation of the X-ray emission might indicate the direction of a
more recent merger event concerning the eastern group (i.e.\ that
centred on the E-peak). We are probably seeing this merging after the
phase of the core passage.  This might explain the offset of the
centre of the X-ray emission, the offset of the DM peak and the small
deviation of the eastern group from the initial NE by E--SW by W
direction. In this case, the eastern group would have shed its gas as
a result of ram pressure at the entry into the main cluster at its SW
by W side and, in fact, we do not see any eastern peak in X-ray
emission (a scenario already suggested by Govoni et al. \cite{gov04}).
Further support for this scenario comes from the fact that emission
line galaxies are preferentially located in the SW cluster region, the
starbursts possibly being induced by the infalling subcluster (e.g.\
Bekki \cite{bek99}).

One expects that cluster mergers will drive shock waves into the
intracluster gas of the two subclusters (see Sarazin \cite{sar02} for
a review). The Mach number of the shock is $\cal M$=${\rm v}_{\rm
s}/c_{\rm s}$, where ${\rm v}_{\rm s}$ is the velocity of the shock
and $c_{\rm s}$ is the sound speed in the preshock gas. In the
stationary regime we can assume that ${\rm v}_{\rm s}$ is the merger
velocity, i.e.\ ${\rm v}_{\rm s} \gtrsim 2300$ \ks taking into account
the projection factor. We obtain $c_{\rm s}\lesssim 1134$ \ks from the
thermal velocity, i.e.\ from the observed $T_{\rm X}$ converted into
velocity using the usual equipartition of energy density between gas
and galaxies (see Sect.~5) and taking into account that the observed
$T_{\rm X}$ might be enhanced with respect to pre-merging values. We
therefore estimate $\cal M$ $\gtrsim 2.0$. According to our merging
scenario, this value of $\cal M$ would make front or bow shocks
possibly detectable in X-ray observations.  However, in the case of
the LOS being very close to the merger axis, a front shock would
expand in an almost parallel direction to the LOS. So because of the
geometry of the problem we can only observe a global enhancement of
the X-ray temperature in the whole cluster.  To date, the
observational picture is inconclusive.  Govoni et al. (\cite{gov04})
found a very hot region in the western cluster region, but, analysing
a slightly larger and more homogeneous amount of data, we find no
significant variation in the X-ray temperature.

As for the comparison with the radio halo feature, its emission is
somewhat displaced towards the E-peak (see Fig.~\ref{figisofote} and
Govoni et al. \cite{gov04}).  This is a suggestive indication in
favour of the connection between cluster mergers and radio halos.
Recent calculations (Cassano \& Brunetti \cite{cas05}) show that a
fraction $\sim$10\% of the thermal-cluster energy may be channeled in
the form of turbulence in case of main (i.e.\ mass ratio$\leq$ 3:1)
cluster mergers and this may power up radio halos.  However, a typical
massive cluster is formed by a series of minor mergers and a minor
merger may still power the particle acceleration process especially if
they happen in clusters which are already dynamically disturbed by
precedent mergers (see Fig.~2 and the time-evolution of the
acceleration coefficient in Fig.~5 in Cassano \& Brunetti
\cite{cas05}). This is probably the case of A773.

\subsection{LSS structure}

We also analyse the large scale structure (LSS) surrounding A773 since
the merging direction was shown to be frequently correlated with the
supercluster environment (e.g.\ Durret et al. \cite{dur98}; Arnaud et
al. \cite{arn00}). We find three Abell clusters within a circle of 2
degrees centred on A773: A782 of richness class $R=2$ and A746 and
A793 of richness class $R=1$. The closest one, A782 at 40.6\arcmm, has
five galaxies with known redshift at 0.20--021 within a circle of
5\arcm as recovered by using NED\footnote{The NASA/IPAC Extragalactic
Database (NED) is operated by the Jet Propulsion Laboratory,
California Institute of Technology, under contract with the National
Aeronautics and Space Administration.}; we therefore assume a cluster
redshift of 0.21. Assuming the respective redshifts, the distance of
A782 to A773 is $\sim 40$ \h thus comparable to the typical sizes of
superclusters (e.g.\ Batuski \& Burns \cite{bat85}; Zucca et
al. \cite{zuc93}).  The direction of A782 is indicated in Figure
\ref{figisofote}: the result is that the direction of the elongation
of X-ray emission and the direction indicated by the eastern group are
$\sim 10$--20 degrees different with the axis A773/A782.  Therefore,
we speculate that A773 and A782 lie along the same LSS filament where
the eastern group also originated. In this case, to take into account
the geometry of the problem and the relative velocity difference of
the eastern group and the main system, the eastern group lies in front
of A773 and has already reached the turnaround point, with a second
centre passage no incipient.

\section{Summary and conclusions}

We present the results of a dynamical analysis of the rich, X-ray
luminous and hot cluster of galaxies A773 containing a diffuse radio
halo.

Our analysis is based on new redshift data for 142 galaxies, measured
from 107 spectra obtained at the TNG and 37 spectra recovered from the
CFHT archive in a cluster region within a radius of $\sim1.8$ \h from
the cluster centre. We also use new photometric data obtained at the
INT in a field larger than 30\arcmm$\times$30\arcm.

We select 100 cluster members with $62250<{\rm v}<68850$ \kss and 
compute a global LOS velocity dispersion of galaxies, 
$\sigma_{\rm v}=1394_{-68}^{+84}$ \kss.


The 2D distribution shows two significant peaks separated by
$\sim$2\arcm along the EW direction with the main, western one located
close to the position of the two dominant galaxies and the X-ray peak.
The velocity distribution of cluster galaxies shows two peaks at ${\rm
v}\sim$65000 and $\sim$67500 \kss, corresponding to the velocities of
the two dominant galaxies.  The low velocity structure has high
velocity dispersion -- $\sigma_{\rm v}=800-1100$ \ks -- and its
galaxies are centred on the western 2D peak.  The high velocity
structure has intermediate velocity dispersion -- $\sigma_{\rm v}\sim
500$ \ks -- and is characterized by a complex 2D structure with a
component centred on the western 2D peak and a component centred on
the eastern 2D peak, these components probably being two independent
groups.

Other evidence that A773 is dynamically disturbed comes from 1) the
presence of two X-ray peaks in the X-ray emission in the central
cluster region, offset with respect to possible optical counterparts;
2) the elongation of X-ray emission (PA$\sim$80 degrees) along the NE
by E--SW by W direction, while the two dominant galaxies indicate a
N--S direction; 3) the preferential SW location of emission line
galaxies.

According to our likely scenario the high velocity group surrounding
the high velocity dominant galaxies traces an old merger of the main
cluster with a group along the N--S direction, while the high velocity
group surrounding the eastern 2D peak of galaxies indicates a more
recent merger, but already enough advanced, along the NE by E-SW by W
direction.  Acting on an already dynamically disturbed cluster, this
second merger event would be the likely cause of a very complex
observational picture and of the radio halo, which is somewhat
displaced towards the eastern group.

For the whole galaxy complex we estimate a virial mass $M=1.2$--2.7
\mqui and $M(<1$ \hh)=(5.9--11.1)\mquaa, values comparable to those of
other rich clusters at nearby or moderate redshifts.

\begin{acknowledgements}
We would like to thank Luigina Feretti for many enlightening
discussions and for the VLA radio image she kindly provided us.  We
also thank Andrea Biviano, Dario Fadda and Federica Govoni for useful
discussions. We thank the referee, Florence Durret, for her useful
suggestions.

 This publication is based on observations made on the island of La
Palma with the Italian Telescopio Nazionale Galileo (TNG), operated by
the Fundaci\'on Galileo Galilei -- INAF (Istituto Nazionale di
Astrofisica), and with the Isaac Newton Telescope (INT), operated by
the Isaac Newton Group (ING), in the Spanish Observatorio of the Roque
de Los Muchachos of the Instituto de Astrofisica de Canarias.

 This publication also makes use of data obtained from the Chandra
data archive at the NASA Chandra X-ray center
(http://cxc.harvard.edu/cda/) and of data accessed as Guest User at
the Canadian Astronomy Data Center, which is operated by the Dominion
Astrophysical Observatory for the National Research Council of
Canada's Herzberg Institute of Astrophysics
(http://cadcwww.dao.nrc.ca/cfht/cfht.html).

 This research has made use of the NASA/IPAC Extragalactic Database
(NED), which is operated by the Jet Propulsion Laboratory, California
Institute of Technology, under contract with the National Aeronautics
and Space Administration.

This work was partially supported by a grant from the Istituto
Nazionale di Astrofisica (INAF, grant PRIN-INAF2005 CRA ref number
1.06.08.05).
\end{acknowledgements}

\clearpage


\begin{table}[!ht]
        \caption[]{Velocity catalog of 142 spectroscopically measured galaxies. In Column~1, IDs in italics indicate non--cluster galaxies.
IDs 59 and 60 indicate the two dominant galaxies (D1 and D2, respectively).}
         \label{catalogue}
              $$ 
           \begin{array}{r c c c c r c}
            \hline
            \noalign{\smallskip}
            \hline
            \noalign{\smallskip}

\mathrm{ID} & \mathrm{\alpha},\mathrm{\delta}\,(\mathrm{J}2000)  & \mathrm{B} & \mathrm{R}  & \mathrm{v} & \mathrm{\Delta}\mathrm{v} & \mathrm{SC} \\
  & 09^h      , +51^o    &  &  &\,\,\,\,\,\,\,\mathrm{(\,km}&\mathrm{s^{-1}\,)}\,\,\,&  \\
            \hline
            \noalign{\smallskip}   
\textit{1}  & 17\ 00.60 , 42\ 52.6 & 19.77 & 18.36 &  34762 &   70 &  \rm{e(c)} \\    
2           & 17\ 06.58 , 44\ 07.1 & 20.54 & 18.86 &  64233 &  105 &        \\    
\textit{3}  & 17\ 13.85 , 43\ 17.0 & 22.05 & 19.88 & 132256 &   93 &        \\   
4           & 17\ 14.04 , 43\ 38.6 & 21.24 & 19.29 &  63970 &   99 &  \rm{k+a}  \\   
5           & 17\ 14.50 , 43\ 28.2 & 21.73 & 20.46 &  65568 &  123 &        \\   
\textit{6}  & 17\ 17.21 , 44\ 43.4 & 22.72 & 19.95 & 132667 &   76 &        \\   
7           & 17\ 19.80 , 38\ 48.8 & 21.38 & 19.39 &  68615 &  147 &        \\   
8           & 17\ 20.86 , 38\ 45.2 & 19.63 & 18.19 &  68559 &   94 &  \rm{e(a)} \\   
9           & 17\ 22.99 , 41\ 18.6 & 22.50 & 20.17 &  64925 &  236 &        \\   
10          & 17\ 23.14 , 40\ 51.6 & 22.46 & 20.09 &  65517 &  144 &        \\   
11          & 17\ 24.05 , 38\ 15.4 & 20.67 & 19.29 &  64503 &   87 &  \rm{e(a?)} \\    
12          & 17\ 26.59 , 43\ 40.4 & 21.00 & 19.47 &  62362 &  124 &        \\    
\textit{13} & 17\ 27.86 , 37\ 33.2 & 20.53 & 18.27 &  94246 &   76 &        \\    
14          & 17\ 28.73 , 42\ 23.0 & 21.30 & 19.28 &  62282 &   58 &    \rm{k}  \\    
15          & 17\ 28.75 , 43\ 50.2 & 21.39 & 20.33 &  63859 &   86 &   \rm{e(b)} \\     
\textit{16} & 17\ 29.71 , 42\ 59.4 & 20.93 & 18.76 &  94885 &   76 &   \rm{e(c)} \\     
17          & 17\ 29.78 , 43\ 08.0 & 21.27 & 19.32 &  68655 &  118 &        \\    
18          & 17\ 31.66 , 44\ 43.1 & 21.81 & 20.34 &  68206 &   74 &        \\   
19          & 17\ 32.50 , 42\ 19.9 & 21.77 & 19.25 &  65171 &  111 &  \rm{e(c?)} \\   
20          & 17\ 32.57 , 42\ 22.2 & 22.21 & 20.72 &  64575 &  136 &        \\  
21          & 17\ 34.25 , 41\ 13.2 & 20.64 & 18.50 &  65848 &   69 &     \rm{k}  \\  
22          & 17\ 35.02 , 44\ 32.6 & 22.06 & 19.73 &  67320 &   68 &        \\  
23          & 17\ 35.09 , 41\ 18.6 & 20.97 & 19.13 &  65309 &   86 &     \rm{k}  \\  
24          & 17\ 35.38 , 38\ 39.8 & 20.44 & 18.73 &  63729 &  105 &   \rm{k+a}  \\  
25          & 17\ 35.59 , 41\ 17.9 & 20.66 & 18.98 &  63067 &   99 &        \\   
\textit{26} & 17\ 36.10 , 44\ 18.2 & 20.75 & 19.44 &  58492 &   81 &        \\  
27          & 17\ 36.29 , 45\ 27.0 & 21.89 & 19.73 &  67997 &   87 &        \\  
\textit{28} & 17\ 37.08 , 42\ 53.6 & 22.26 & 19.87 &  93910 &   73 &        \\  
29          & 17\ 37.37 , 42\ 36.4 & 22.85 & 20.73 &  64940 &  212 &        \\  
30          & 17\ 37.99 , 44\ 17.5 & 20.22 & 18.08 &  66016 &   51 &     \rm{k}  \\   
31          & 17\ 38.66 , 43\ 25.7 & 20.60 & 19.30 &  64889 &   85 &  \rm{e(a?)} \\  
\textit{32} & 17\ 38.74 , 46\ 58.4 & 19.67 & 18.84 &  20854 &  114 &  \rm{e(a?)} \\  
33          & 17\ 39.22 , 46\ 53.8 & 20.93 & 18.80 &  64216 &  118 &     \rm{k}  \\  
34          & 17\ 39.48 , 41\ 49.6 & 21.97 & 20.20 &  67577 &   70 &        \\  
35          & 17\ 40.73 , 38\ 18.6 & 19.95 & 18.72 &  67396 &   86 &        \\  
36          & 17\ 40.80 , 41\ 31.6 & 22.54 & 20.13 &  65289 &  206 &        \\  
\textit{37} & 17\ 41.06 , 43\ 58.4 & 22.19 & 20.68 & 124666 &   24 &  \rm{e(c?)} \\  
\textit{38} & 17\ 42.79 , 43\ 41.2 & 20.90 & 19.77 &  55811 &   93 &  \rm{e(c)}  \\  
39          & 17\ 44.04 , 40\ 21.0 & 22.51 & 20.52 &  67468 &  148 &  \rm{e(c?)} \\  
40          & 17\ 44.66 , 43\ 26.4 & 21.73 & 19.63 &  65758 &   70 &     \rm{k}  \\  
41          & 17\ 45.72 , 42\ 55.1 & 21.66 & 19.52 &  64248 &   72 &     \rm{k}  \\  
42          & 17\ 46.32 , 43\ 13.1 & 21.26 & 18.90 &  65547 &   50 &     \rm{k}  \\  
43          & 17\ 46.85 , 45\ 29.5 & 20.74 & 18.64 &  65902 &   54 &     \rm{k}  \\  
44          & 17\ 47.14 , 40\ 59.2 & 21.26 & 19.03 &  65661 &  106 &     \rm{k}  \\  
45          & 17\ 48.02 , 46\ 35.4 & 20.30 & 18.65 &  66653 &  112 &     \rm{k}  \\  
46          & 17\ 48.22 , 46\ 13.1 & 20.85 & 18.76 &  66848 &   60 &     \rm{k}  \\  
47          & 17\ 48.62 , 43\ 43.7 & 22.07 & 20.81 &  64023 &   90 &        \\  
\textit{48} & 17\ 48.82 , 40\ 59.9 & 19.90 & 18.78 &  44429 &   81 &   \rm{e(a)} \\  
\textit{49} & 17\ 48.89 , 44\ 53.5 & 20.05 & 18.73 &  55538 &   92 &   \rm{e(a)} \\  
50          & 17\ 49.46 , 43\ 08.4 & 21.69 & 19.18 &  68830 &   76 &     \rm{k}  \\ 
               \noalign{\smallskip}			    
            \hline					    
            \noalign{\smallskip}			    
            \hline					    
         \end{array}
     $$ 
         \end{table}
\addtocounter{table}{-1}
\begin{table}[!ht]
          \caption[ ]{Continued.}
     $$ 
           \begin{array}{r c c c c r c}
            \hline
            \noalign{\smallskip}
            \hline
            \noalign{\smallskip}

\mathrm{ID} & \mathrm{\alpha},\mathrm{\delta}\,(\mathrm{J}2000)  & \mathrm{B} & \mathrm{R}  & \mathrm{v} & \mathrm{\Delta}\mathrm{v} & \mathrm{SC} \\
  & 09^h      , +51^o    &  &  &\,\,\,\,\,\,\,\mathrm{(\,km}&\mathrm{s^{-1}\,)}\,\,\,&  \\
            \hline
            \noalign{\smallskip} 

51          & 17\ 49.66 , 41\ 04.9 & 19.88 & 17.64 &  65088 &   62 &     \rm{k}  \\  
52          & 17\ 49.82 , 43\ 51.0 & 22.67 & 21.27 &  67857 &   58 &  \rm{e(c?)} \\  
53          & 17\ 50.54 , 43\ 26.4 & 20.95 & 18.63 &  63900 &   38 &     \rm{k}  \\  
54          & 17\ 51.00 , 43\ 16.0 & 22.27 & 19.68 &  63696 &   64 &        \\  
\textit{55} & 17\ 51.96 , 41\ 33.7 & 20.13 & 19.04 & 114971 &   75 &   \rm{e(a)} \\  
56          & 17\ 52.56 , 43\ 50.2 & 20.84 & 18.31 &  64849 &   46 &     \rm{k}  \\  
57          & 17\ 53.09 , 42\ 55.4 & 20.33 & 18.05 &  67364 &   72 &     \rm{k}  \\  
58          & 17\ 53.16 , 44\ 08.9 & 20.81 & 18.59 &  64194 &   74 &     \rm{k}  \\  
59          & 17\ 53.26 , 43\ 36.5 & 18.82 & 16.49 &  65131 &   50 &     \rm{k}  \\  
60          & 17\ 53.33 , 44\ 00.6 & 18.93 & 16.73 &  67765 &   56 &     
\rm{k}  \\ 
61          & 17\ 53.38 , 44\ 24.0 & 21.40 & 19.05 &  65152 &  124 &     \rm{k}  \\  
62          & 17\ 54.17 , 43\ 27.5 & 21.26 & 18.87 &  67578 &  126 &     \rm{k}  \\   
63          & 17\ 54.17 , 43\ 55.6 & 19.90 & 17.83 &  66477 &   28 &     \rm{k}  \\   
64          & 17\ 54.24 , 44\ 52.8 & 20.44 & 18.58 &  68485 &   66 &     \rm{k}  \\  
\textit{65} & 17\ 54.82 , 42\ 46.8 & 22.46 & 19.99 &  79178 &  158 &        \\ 
\textit{66} & 17\ 54.84 , 47\ 03.8 & 18.84 & 17.77 &  21910 &  116 &   \rm{e(a)} \\  
67          & 17\ 55.54 , 43\ 54.1 & 20.77 & 18.49 &  66543 &   30 &     \rm{k}  \\  
\textit{68} & 17\ 55.97 , 41\ 51.0 & 20.85 & 19.24 &  55993 &   87 &   \rm{e(c)} \\  
69          & 17\ 56.76 , 43\ 28.6 & 21.13 & 18.89 &  64437 &   57 &     \rm{k}  \\ 
70          & 17\ 57.10 , 46\ 10.9 & 21.20 & 19.00 &  68250 &   50 &     \rm{k}  \\ 
71          & 17\ 57.82 , 43\ 27.4 & 23.41 & 21.39 &  66885 &  130 &        \\  
72          & 17\ 58.20 , 41\ 35.5 & 21.01 & 18.59 &  67420 &   39 &     \rm{k}  \\  
73          & 17\ 59.76 , 40\ 03.7 & 21.76 & 19.54 &  67738 &   64 &     \rm{k}  \\  
74          & 18\ 00.14 , 46\ 23.2 & 21.24 & 19.12 &  68085 &   51 &     \rm{k}  \\  
75          & 18\ 00.24 , 44\ 11.0 & 20.89 & 19.02 &  62377 &   46 &   \rm{k+a}  \\  
76          & 18\ 00.60 , 44\ 07.4 & 21.30 & 18.86 &  64978 &   33 &     \rm{k}  \\  
77          & 18\ 01.27 , 44\ 13.6 & 21.58 & 19.15 &  68682 &   69 &     \rm{k}  \\  
78          & 18\ 01.37 , 42\ 25.9 & 21.54 & 19.24 &  66377 &   69 &     \rm{k}  \\  
79          & 18\ 01.75 , 40\ 34.0 & 21.27 & 18.85 &  64829 &   72 &     \rm{k}  \\  
80          & 18\ 02.86 , 40\ 07.3 & 21.58 & 19.55 &  63748 &   87 &     \rm{k}  \\  
81          & 18\ 02.86 , 41\ 40.6 & 20.27 & 18.78 &  68358 &   44 &   \rm{a+k}  \\  
82          & 18\ 03.46 , 43\ 51.2 & 20.20 & 17.98 &  67742 &   23 &     \rm{k}  \\  
83          & 18\ 03.55 , 43\ 33.2 & 19.95 & 17.77 &  67969 &   54 &     \rm{k}  \\  
84          & 18\ 03.77 , 43\ 28.2 & 21.29 & 19.24 &  66659 &  166 &        \\  
85          & 18\ 04.01 , 42\ 54.0 & 21.50 & 19.41 &  65774 &   66 &     \rm{k}  \\  
86          & 18\ 04.10 , 43\ 12.7 & 21.05 & 18.76 &  64671 &   46 &     \rm{k}  \\  
87          & 18\ 04.42 , 43\ 44.4 & 19.59 & 17.53 &  67673 &   36 &     \rm{k}  \\  
88          & 18\ 05.06 , 43\ 49.8 & 20.04 & 17.98 &  67047 &   24 &     \rm{k}  \\   
89          & 18\ 05.38 , 41\ 04.2 & 21.80 & 20.04 &  64928 &  112 &   \rm{e(a)} \\   
90          & 18\ 05.57 , 46\ 53.4 & 20.50 & 18.51 &  67098 &   52 &   \rm{k+a}  \\   
\textit{91} & 18\ 05.64 , 39\ 59.0 & 21.46 & 19.38 &  12022 &   72 &        \\  
92          & 18\ 05.71 , 45\ 18.0 & 20.50 & 18.52 &  66287 &   66 &   \rm{e(c)} \\  
93          & 18\ 05.78 , 43\ 40.1 & 20.22 & 18.20 &  65414 &   24 &     \rm{k}  \\  
94          & 18\ 06.22 , 42\ 00.0 & 20.90 & 18.91 &  66502 &   60 &     \rm{k}  \\  
95          & 18\ 07.08 , 44\ 17.2 & 20.51 & 18.23 &  65081 &   27 &     \rm{k}  \\  
\textit{96} & 18\ 07.27 , 49\ 13.8 & 19.56 & 18.44 &  44491 &   88 &   \rm{e(b)} \\  
97          & 18\ 08.57 , 43\ 42.6 & 21.16 & 18.98 &  67860 &   28 &     \rm{k}  \\  
\textit{98} & 18\ 08.57 , 48\ 51.5 & 20.36 & 19.28 &  44498 &   70 &   \rm{e(c)} \\  
99          & 18\ 08.64 , 48\ 11.3 &  0.0  &  0.0  &  65850 &   70 &     \rm{k}  \\  
100         & 18\ 09.79 , 41\ 48.1 & 20.36 & 18.70 &  67183 &   69 &   \rm{e(a)} \\ 
                \noalign{\smallskip}			    
            \hline					    
            \noalign{\smallskip}			    
            \hline					    
         \end{array}
     $$ 
         \end{table}
\addtocounter{table}{-1}
\begin{table}[!ht]
          \caption[ ]{Continued.}
     $$ 
           \begin{array}{r c c c c r c}
            \hline
            \noalign{\smallskip}
            \hline
            \noalign{\smallskip}

\mathrm{ID} & \mathrm{\alpha},\mathrm{\delta}\,(\mathrm{J}2000)  & \mathrm{B} & \mathrm{R}  & \mathrm{v} & \mathrm{\Delta}\mathrm{v} & \mathrm{SC} \\
  & 09^h      , +51^o    &  &  &\,\,\,\,\,\,\,\mathrm{(\,km}&\mathrm{s^{-1}\,)}\,\,\,&  \\
            \hline
            \noalign{\smallskip} 
101         & 18\ 10.13 , 45\ 18.0 & 21.29 & 19.12 &  65867 &  147 &        \\  
102         & 18\ 10.54 , 42\ 26.3 & 20.80 & 18.79 &  66845 &   24 &     \rm{k}  \\ 
103         & 18\ 10.61 , 43\ 14.5 & 21.33 & 19.41 &  67062 &   69 &     \rm{k}  \\  
\textit{104}& 18\ 11.45 , 43\ 44.8 & 20.47 & 19.60 &  55929 &   80 &  \rm{e(b)}  \\  
\textit{105}& 18\ 12.37 , 47\ 48.1 & 21.37 & 19.81 & 101181 &   88 &  \rm{e(c?)} \\  
106         & 18\ 12.98 , 42\ 49.0 & 21.57 & 19.90 &  63437 &   96 &        \\  
107         & 18\ 14.38 , 46\ 43.7 & 20.83 & 19.04 &  63833 &   84 &    \rm{k+a} \\  
\textit{108}& 18\ 14.54 , 43\ 20.9 & 20.27 & 18.10 &  55650 &   34 &     \rm{k}  \\  
109         & 18\ 14.81 , 43\ 49.4 & 21.38 & 19.22 &  64637 &   69 &     \rm{k}  \\  
\textit{110}& 18\ 15.17 , 43\ 21.8 & 19.77 & 17.74 &  55648 &   60 &   \rm{e(a)} \\  
\textit{111}& 18\ 15.84 , 40\ 44.4 & 20.32 & 18.87 &  55650 &   84 &     \rm{k}  \\   
\textit{112}& 18\ 15.98 , 45\ 33.5 & 20.77 & 19.23 &  47297 &  102 &        \\ 
\textit{113}& 18\ 17.04 , 42\ 50.0 & 20.07 & 18.04 &  55700 &   38 &     \rm{k}  \\  
\textit{114}& 18\ 17.11 , 43\ 28.2 & 22.02 & 19.67 &  78329 &   64 &        \\  
115         & 18\ 18.34 , 47\ 53.5 & 21.75 & 19.39 &  64910 &   60 &     \rm{k}  \\  
\textit{116}& 18\ 18.55 , 43\ 49.0 & 20.23 & 17.59 &  86619 &   25 &     \rm{k}  \\  
117         & 18\ 19.66 , 42\ 48.2 & 22.30 & 19.84 &  65556 &   36 &     \rm{k}  \\   
118         & 18\ 21.00 , 43\ 05.5 & 20.67 & 18.38 &  67354 &   35 &     \rm{k}  \\   
119         & 18\ 22.13 , 47\ 56.0 & 21.81 & 19.84 &  64686 &  117 &     \rm{k}  \\  
\textit{120}& 18\ 22.63 , 43\ 09.5 & 21.79 & 20.76 &  44170 &  159 &   \rm{e(b)} \\  
121         & 18\ 23.18 , 48\ 33.1 & 20.79 & 18.47 &  64095 &   57 &    \rm{k}  \\   
122         & 18\ 23.42 , 45\ 22.0 & 21.05 & 18.84 &  64388 &   69 &     \rm{k}  \\   
\textit{123}& 18\ 23.47 , 45\ 56.9 & 21.47 & 19.25 &  74424 &   60 &     \rm{k}  \\   
124         & 18\ 23.74 , 43\ 30.4 & 22.73 & 20.62 &  67352 &   50 &        \\   
\textit{125}& 18\ 23.77 , 47\ 24.9 & 20.20 & 19.38 &  58662 &  141 &   \rm{e(b)} \\   
126         & 18\ 23.93 , 47\ 19.0 & 21.15 & 19.34 &  63302 &   86 &     \rm{k}  \\   
127         & 18\ 24.57 , 47\ 54.5 & 21.50 & 20.20 &  65857 &   92 &     \rm{k}  \\   
\textit{128}& 18\ 24.72 , 42\ 01.4 & 21.84 & 20.19 & 102297 &   47 &     \rm{k}  \\   
\textit{129}& 18\ 25.10 , 49\ 09.1 & 20.78 & 19.43 & 128313 &   96 &  \rm{e(c?)} \\   
\textit{130}& 18\ 27.72 , 41\ 13.2 & 22.15 & 20.43 & 130562 &   72 &        \\   
\textit{131}& 18\ 27.72 , 46\ 05.9 & 20.52 & 19.52 &  21763 &  130 &        \\   
132         & 18\ 27.73 , 44\ 27.7 & 21.36 & 19.38 &  65252 &  110 &     \rm{k}  \\   
133         & 18\ 29.57 , 40\ 42.6 & 21.57 & 19.48 &  67741 &   33 &     \rm{k}  \\   
\textit{134}& 18\ 30.21 , 47\ 21.9 & 22.38 & 20.21 & 138204 &   74 &        \\    
\textit{135}& 18\ 30.46 , 44\ 23.6 & 20.86 & 19.64 &  52800 &   90 &   \rm{e(c)} \\    
136         & 18\ 31.73 , 42\ 49.3 & 20.60 & 18.51 &  64282 &   66 &     \rm{k}  \\     
\textit{137}& 18\ 35.30 , 45\ 26.6 & 19.77 & 17.60 &  78472 &   56 &    \rm{k+a} \\     
\textit{138}& 18\ 40.66 , 44\ 12.1 & 20.34 & 18.34 &  52942 &   64 &     \rm{k}  \\    
\textit{139}& 18\ 41.33 , 43\ 10.6 & 21.11 & 20.13 &  29057 &  104 &   \rm{e(c)} \\     
140         & 18\ 44.45 , 44\ 14.3 & 21.78 & 19.71 &  65279 &  122 &        \\     
\textit{141}& 18\ 44.86 , 43\ 09.8 & 19.92 & 18.21 &  52748 &   63 &     \rm{k}  \\     
\textit{142}& 18\ 49.37 , 45\ 19.8 & 20.72 & 19.30 & 113177 &  231 &   \rm{e(a)} \\ 
              \noalign{\smallskip}			    
            \hline					    
            \noalign{\smallskip}			    
            \hline					    
         \end{array}\\
     $$ 
\label{tab1}

\end{table}


\end{document}